\documentclass[twocolumn,prd,showpacs,preprintnumbers,amsmath,amssymb,floatfix]{revtex4}

\usepackage{graphicx}
\usepackage{dcolumn}
\usepackage{bm}

\newcommand{\Slash}[1]{\ooalign{\hfil/\hfil\crcr$#1$}}
\newcommand{\re}{\text{Re}}
\newcommand{\im}{\text{Im}}
\newcommand{\Tr}{\text{Tr}}

\newcommand{\bra}[1]{\langle \, #1 \, |}
\newcommand{\ket}[1]{| \, #1 \, \rangle}
\newcommand{\llim}{\beta_{-}}
\newcommand{\ulim}{\beta_{+}}

\newcommand{\PDG}{Yao:2006px}

\newcommand{\iso}[6]{\mbox{$\left( \begin{array}{cc||c} {#1} & {#2} &
{#3} \\ {#4} & {#5} & {#6} \end{array} \right)$}}

\newcommand{\largeN}[1]{\text{``$#1$''}}

\begin{document}

\preprint{YITP-06-58, RCNP-Th06029}

\title{Study of exotic hadrons in $S$-wave scatterings 
induced by chiral interaction \\
in the flavor symmetric limit}

\author{Tetsuo~Hyodo}
\email{hyodo@yukawa.kyoto-u.ac.jp}
\affiliation{%
Yukawa Institute for Theoretical Physics,
Kyoto University, Kyoto 606--8502, Japan
}%
 
\author{Daisuke~Jido}
\affiliation{%
Yukawa Institute for Theoretical Physics,
Kyoto University, Kyoto 606--8502, Japan
}%
 
\author{Atsushi~Hosaka}%
\affiliation{%
Research Center for Nuclear Physics (RCNP),
Ibaraki, Osaka 567-0047, Japan
}%

\date{\today}
\begin{abstract}
    We study $s$-wave bound states of a hadron and a light pseudoscalar 
    meson induced by the Weinberg-Tomozawa (WT) interaction in the flavor 
    SU(3) symmetric limit. The WT interaction is a driving force to generate
    quasibound states dynamically in the chiral unitary approaches.
    The strength and sign of the WT interaction are determined only by the 
    group theoretical structure of the target hadrons, and we present a 
    general expression of the strengths for the flavor SU(3) case. We show 
    that, for the channels which are more exotic than the target, the
    interaction is repulsive in most cases, and the strength of the 
    attractive 
    interaction is universal for any possible target states. We demonstrate 
    that the attractive coupling is not strong enough to generate an exotic 
    state from the physically known masses of target hadrons.
    In addition, we also find a nontrivial $N_c$ dependence of the coupling 
    strengths. We show that the channels which are attractive at $N_c=3$ 
    changes into repulsive ones for large $N_c$, and, therefore, no 
    attractive interaction exists in exotic channels in the large-$N_c$ 
    limit. 
\end{abstract}




\maketitle

\section{Introduction}

The hadrons observed in experiments are classified by their spin and flavor 
quantum numbers~\cite{\PDG}. The flavor quantum numbers are ordinarily 
composed of the valence quark contents of $\bar q q$ for mesons and $qqq$ 
for baryons. The mesons and baryons with flavor that cannot be achieved
by $\bar qq$ or $qqq$ are called exotic hadrons, and must have more valence 
quarks than the ordinary hadrons. So far, the existence of the exotic state 
has not been clearly established. The evidence for the pentaquark 
$\Theta^{+}$ at LEPS/SPring8~\cite{Nakano:2003qx} is one of the possible 
candidates, but the experimental confirmation of the existence of 
$\Theta^{+}$ is still controversial~\cite{Hicks:2005gp}. The structure of 
the $X(3872)$~\cite{Choi:2003ue} and $D_{s}(2317)$~\cite{Aubert:2003fg} is 
also found to be quite different from a simple $\bar q q$ state, although 
their flavor quantum numbers can be achieved by $\bar q q$. 

The definitive answer for whether the exotics exist or not will be given by 
future experiments, but it is highly nontrivial that such states are almost 
completely absent in the hadron spectrum. The fundamental theory of the 
strong interaction, QCD, does not forbid the exotic states. There is no 
simple selection rule to exclude such states in effective models describing 
the ordinary hadrons well. In this context, we have studied the nature of 
the exotic states as hadronic molecule states in Ref.~\cite{Hyodo:2006yk} 
based on the $s$-wave dynamics with the underlying chiral symmetry in QCD.

In recent years, there have been remarkable developments in theoretical 
studies of hadron spectroscopy. Some of the baryon resonances are well 
described as meson-baryon quasibound states in chiral unitary 
approaches~\cite{Kaiser:1995eg,Oset:1998it,Krippa:1998us,Oller:2000fj,
Lutz:2001yb}. These resonances are dynamically generated in meson-baryon 
scattering formulated by unitarization of low energy interactions governed 
by chiral symmetry. This approach is along the same lines as the 
coupled-channel analysis developed in the 1960's~\cite{Dalitz:1960du,
PR155.1649}, and recent studies revealed the novel structure of the 
$\Lambda(1405)$, namely, the existence of two poles in the region of 
$\Lambda(1405)$~\cite{Jido:2003cb,Hyodo:2003jw,Magas:2005vu}. The method was
also applied to the scattering of the pseudoscalar mesons with the decuplet 
baryons of $J^P=3/2^+$, leading to the dynamical generation of the 
$J^P=3/2^-$ resonances~\cite{Kolomeitsev:2003kt,Sarkar:2004jh}. Applications
to the charmed baryon sector~\cite{Lutz:2003jw} and the heavy meson 
sector~\cite{Kolomeitsev:2003ac,Guo:2006fu,Guo:2006rp} were also performed,
in fair agreement with the experimental observation. In these studies, an 
important ingredient to generate $s$-wave resonances dynamically was the 
Weinberg-Tomozawa (WT) interaction of hadrons and the Nambu-Goldstone (NG) 
bosons~\cite{Weinberg:1966kf,Tomozawa:1966jm}.

The WT term has unique features in chiral dynamics. The kinematical 
structure of the coupling is fixed as an $s$-wave interaction between the 
target hadron and NG boson system with linear dependence of the NG boson 
energy. Also, its sign and strength are universally determined by chiral 
symmetry, once one fixes the flavor structure of the hadron. This is because
the interaction can be derived from the current algebra, even without using 
the chiral Lagrangians. In addition, the WT term is the leading order term 
of the chiral expansion, and therefore it gives the dominant contribution at
low energy. Thus, the low energy hadron interaction in an $s$-wave is 
governed by the WT term, which is model independent, as far as we respect 
chiral symmetry.

In this paper, we would like to present a detailed analysis of 
Ref.~\cite{Hyodo:2006yk}, studying the mechanism of the generation of 
resonances of the NG boson and a target hadron in a simplified and essential
version of the chiral unitary approaches, paying attention to the flavor 
exotic states. The dynamically generated state in this approach is 
considered as a quasibound state of the NG boson and the target hadron. 
Since such a generated state has additional $\bar{q}q$ components, it is 
natural to consider exotic states in this approach. To start with, we take 
flavor SU(3) symmetric limit for simplicity. In the SU(3) limit, the 
complicated coupled-channel equations reduce to a set of independent 
single-channel problems. In Refs.~\cite{Jido:2003cb,Garcia-Recio:2003ks}, 
some resonances in the physical world of SU(3) breaking were shown to turn 
into bound states in the limit of SU(3) symmetry. The existence of the bound
states in the SU(3) limit was also confirmed in different 
channels~\cite{Kolomeitsev:2003kt,Sarkar:2004jh,Lutz:2003jw,
Kolomeitsev:2003ac}. Therefore, in this limit, our task is to search for 
bound state poles in the scattering amplitude on the real axis below 
threshold, which are considered to be the origin of the physical resonances 
with SU(3) breaking.

This paper is organized as follows. In Sec.~\ref{sec:ChU}, we review the 
formulation of the chiral unitary approach and show how the physical 
resonances are dynamically generated. In Sec.~\ref{sec:WTterm}, we focus 
on the WT term, that is, the kernel interaction of the chiral unitary 
approach. Based on a group theoretical argument, we present the general 
formula of the coupling strength of the WT term. Its expressions give us the
important consequence that the interaction in exotic channels is repulsive 
in most cases, and the possible strength of the attraction is independent of
channels. The behavior of the coupling strength in the large-$N_c$ limit is 
also discussed. In Sec.~\ref{sec:Boundstate}, we derive the condition to 
generate a bound state, based on the general principle of the scattering 
theory. Presenting numerical analysis with physical masses and coupling 
constants, we show under which condition the bound states are generated in 
the exotic channels. The last section is devoted to a summary.

\section{Chiral unitary approach}\label{sec:ChU}

In this section, we present the formulation of the chiral unitary approach
that describes the scattering of a NG boson with a target hadron. There are 
two important ingredients, chiral symmetry of the interaction and the 
unitarity of the scattering amplitude. The scattering amplitude is 
nonperturbatively constructed so as to maintain the unitarity and to match 
the kernel interaction with the chiral perturbation theory in the low energy
limit. The resonances can be dynamically generated in the resulting 
amplitude which agrees well with experimental data for various target 
hadrons. 

\subsection{Kernel interaction}

Here we introduce the Weinberg-Tomozawa interaction~\cite{Weinberg:1966kf,
Tomozawa:1966jm}, focusing on the kinematic structure. To start with, let 
us take an example of the scattering of the pseudoscalar octet mesons (the 
NG bosons of three flavor chiral symmetry) with the ground state octet 
baryons. In the chiral perturbation theory, the WT interaction is derived as
the leading order term of the chiral expansion in the covariant derivative 
of the kinetic term,
\begin{align}
    \Tr \left(\bar{B}i\Slash{D}B\right) 
    =\Tr \left(\bar{B}i\Slash{\partial}B\right)
    +\Tr \left(\bar{B}i\gamma^{\mu}\frac{1}{4f^2}[
    \Phi\overleftrightarrow{\partial_{\mu}}\Phi,B] \right)
    +\dots ,
    \label{eq:WTterm}
\end{align}
with the meson decay constant $f$ in the SU(3) limit and the baryon octet 
($B$) and the pseudoscalar meson octet ($\Phi$) fields~\cite{Bernard:1995dp,
Ecker:1995gg,Pich:1995bw,Hyodo:2003qa}. This term provides the interaction 
of the meson-baryon scattering
\begin{equation}
    V_{ij}^{(WT)}
    =-\frac{1}{4f^2}C_{ij}
    \bar{u}(\Slash{k}_i+\Slash{k}_j)u ,
    \nonumber
\end{equation}
where $i,j$ denote the meson-baryon channels in the final and initial 
states, $k_i$ and $k_{j}$ are the momenta of the meson in channels $i$ and 
$j$, respectively, $u$ and $\bar{u}$ are the baryon spinors, and $C_{ij}$ is
the coupling strength matrix in channel space.
 
The sign and strength of the matrix element $C_{ij}$ are determined by the 
flavor symmetry. In general, the matrix $C_{ij}$ has the off-diagonal 
components which are responsible for the transition between two different 
channels. Therefore, solving the scattering equation is a coupled-channel 
problem. On the other hand, $C_{ij}$ becomes a diagonal matrix in the basis 
of the SU(3) irreducible representations, since the interaction is SU(3) 
symmetric. Using the SU(3) Clebsch-Gordan coefficients 
$\langle i,\alpha\rangle $ which relates the particle basis $i$ with the
SU(3) basis $\alpha$ symbolically, the matrix $C_{ij}$ can be transformed 
into 
that in the SU(3) basis as $C_{\alpha\beta} =\sum_{i,j}\langle i,\alpha
\rangle C_{ij}\langle j,\beta\rangle=C_{\alpha}\delta_{\alpha\beta}$, where
$\alpha$ and $\beta$ denote the irreducible representations of 
SU(3)~\cite{Jido:2003cb}. Since this matrix is diagonal in the SU(3) basis,
the coupled-channel equations reduce to a set of independent single-channel
equations. 

In the SU(3) limit, the interaction is written in the nonrelativistic 
reduction as
\begin{align}
    V^{(WT)}_{\alpha \beta}
    \sim &-\frac{\omega}{2f^2}C_{\alpha , T}\delta_{\alpha\beta} ,
    \label{eq:WTint} 
\end{align}
with the energy of the meson $\omega$ and the diagonal matrix 
$C_{\alpha ,T}$ for which we put the index of the target representation $T$ 
for later convenience. 

It is important to note that the structure and strength of the WT term are 
determined by chiral symmetry without introducing additional coupling
constants; the constant $f$ can be determined by the weak decay of the NG 
boson. In addition, the structure of the coupling is universally fixed for 
any target states, since Eq.~\eqref{eq:WTint} is also derived from current 
algebra, without using chiral Lagrangian~\cite{Hyodo:2006yk}. Therefore, 
although Eq.~\eqref{eq:WTterm} is given for the system of the NG boson and 
the octet baryons, we can generalize Eq.~\eqref{eq:WTint} to that of any 
hadron targets. For a meson target, some minor modifications should be made
in Eq.~\eqref{eq:WTint} and the formulas of this section. Nevertheless, we 
find that the expression given in this section can be applied to the meson 
target in the heavy mass approximation. The detailed discussion is given in 
Appendix~\ref{sec:mesoncase}.

\subsection{Unitarization and regularization}

With the WT interaction in the perturbation theory, the scattering amplitude
does not satisfy the unitarity condition. Here we show the way to maintain 
the unitarity~\cite{Oller:1998zr,Oller:2000fj}. It necessarily brings a
regularization procedure for divergent loop integrals~\cite{Lutz:2001yb}. We
again take the example of the meson-baryon scattering.

In the chiral unitary approaches, we impose the unitarity condition for the 
scattering amplitude, based on the N/D method~\cite{Hyodo:2003qa}. Assuming
the elastic unitarity and neglecting the left-hand cut from the crossed 
diagrams, the meson-baryon scattering amplitude reads
\begin{equation}
    t(\sqrt{s})=\frac{1}{1-V(\sqrt{s})G(\sqrt{s})}V(\sqrt{s}) ,
    \label{eq:ChUamp}
\end{equation}
as a function of the center-of-mass energy $\sqrt{s}$. Here $V(\sqrt{s})$ 
denotes the WT interaction \eqref{eq:WTint} where $\omega = (s-M_T^2+m^2 )/ 
(2\sqrt{s})$ with $m$ and $M_T$ being masses of the meson and baryon, 
respectively. The function $G(\sqrt{s})$ is given by the spectral 
representation
\begin{align}
    G(\sqrt{s})
    &=-\tilde{a}(s_0)
    -\frac{1}{2\pi}
    \int_{s^{+}}^{\infty}ds^{\prime}
    \left(
    \frac{\rho(s^{\prime})}{s^{\prime}-s}
    -\frac{\rho(s^{\prime})}{s^{\prime}-s_0}
    \right) ,
    \label{eq:loop_s}  
\end{align}
where $s^{+}=(m+M_T)^2$ is the value of $s$ at the threshold. The parameter 
$\tilde{a}(s_0)$ is the subtraction constant which is not determined within 
the N/D method. The integrated phase space is
\begin{equation}
    \rho(s) = \frac{2M_T\bar{q}}{4\pi \sqrt{s}} ,
  \label{eq:phasespace}
\end{equation}
where $\bar{q}=\lambda^{1/2}(s,M_T^2,m^2)/2\sqrt{s}$ and the K\"allen 
function $\lambda^{1/2}(x,y,z)=x^2+y^2+z^2-2xy-2yz-2zx$.

In principle, the renormalization of the amplitude should be performed so 
as to reproduce some experimental observables. Here, to fix the subtraction 
parameter, we adopt the prescription given in Ref.~\cite{Lutz:2001yb}, which
gives the proper subtraction constant for the $S=-1$ meson-baryon channel as
we will see below. In this scheme, the scattering amplitude is adjusted to 
be the kernel interaction $V$ at $\sqrt s = M_T$ with the regularization 
scale $\mu = M_T$:
\begin{align}
    G(\mu)=&0  \quad \Leftrightarrow \quad t(\mu)=V(\mu) 
    \quad \text{at} \   \mu = M_T .
    \label{eq:regucond} 
\end{align}
This is equivalent to taking the subtraction point at the baryon mass:
\begin{equation}
    s_0=M_T^2 , \quad \tilde{a}(M_T^2)=0 .
    \nonumber
\end{equation}
This condition determines the energy at which the chiral perturbation theory
works. At the same time, this condition guarantees that the present 
amplitude with $s$-channel unitarity coincides with the crossed amplitude 
with $u$-channel unitarity at $\sqrt{s}=M_T$. This prescription was first 
introduced in the $\pi$-$\pi$ 
scattering~\cite{Igi:1998gn}, and was applied to the meson-baryon scattering
in Ref.~\cite{Lutz:2001yb}.

It is instructive to show that the algebraic form of Eq.~\eqref{eq:ChUamp} 
derived in the N/D method was proved to be equivalent to the solution of the
on-shell factorized Bethe-Salpeter integral equation~\cite{Oller:2000fj} if 
one identifies the $G$ function~\eqref{eq:loop_s} with the meson-baryon loop
function,
\begin{align}
    G(\sqrt{s})
    =&
    i\int\frac{d^{4}q}{(2\pi)^{4}}
    \frac{2M_T}{(P-q)^{2}-M_T^{2}+i\epsilon}
    \frac{1}{q^{2}-m^{2}+i\epsilon}
    \nonumber \\
    =&\frac{2M_T}{(4\pi)^{2}}
    \Bigl\{a(\mu)+\ln\frac{M_T^{2}}{\mu^{2}}
    +\frac{m^{2}-M_T^{2}+s}{2s}\ln\frac{m^{2}}{M_T^{2}}
    \nonumber\\
    &+\frac{\bar{q}}{\sqrt{s}}
    [\ln(s-(M_T^{2}-m^{2})+2\sqrt{s}\bar{q})
    \nonumber\\
    &
    +\ln(s+(M_T^{2}-m^{2})+2\sqrt{s}\bar{q}) 
    \nonumber\\
    &
    -\ln(-s+(M_T^{2}-m^{2})+2\sqrt{s}\bar{q})\nonumber\\
    &
    -\ln(-s-(M_T^{2}-m^{2})+2\sqrt{s}\bar{q})
    ]\Bigr\} ,
    \nonumber 
\end{align}
where $P^{\mu}=(\sqrt{s},\bm{0})$, $\mu$ and $a(\mu)$ are the regularization
scale and subtraction constant, which correspond to $s_0$ and 
$\tilde{a}(s_0)$ in Eq.~\eqref{eq:loop_s}. This loop function is exactly 
the same as the dispersion integral~\eqref{eq:loop_s} with the phase 
space~\eqref{eq:phasespace} up to constants.

The renormalization condition~\eqref{eq:regucond} is suitable for the 
present purpose, since the regularization parameters $a$ and $\mu$ are 
systematically fixed by the baryon and meson masses without other 
experimental inputs:
\begin{align}
    \mu =& M_T , \nonumber\\ 
    a(M_T)
    =&
    -\Bigl\{\frac{m^{2}}{2M_T^2}\ln\frac{m^{2}}{M_T^{2}}
    +\frac{m\sqrt{m^2-4M_T^2}}{2M_T^2} \nonumber \\
    &
    \times [\ln(m^{2}+m\sqrt{m^2-4M_T^2})\nonumber \\
    &
    +\ln(2M_T^{2}-m^{2}+m\sqrt{m^2-4M_T^2}) \nonumber \\
    &
    -\ln(-m^{2}+m\sqrt{m^2-4M_T^2})\nonumber \\
    &
    -\ln(-2M_T^{2}+m^{2}+m\sqrt{m^2-4M_T^2})
    ]\Bigr\} \label{eq:adet} .
\end{align}
This is not the unique choice of the subtraction parameter. Generally, the 
subtraction constant is fixed at the renormalization point by experimental 
inputs, such as the scattering length. One can also determine $a(\mu)$ so as
to reproduce the observed threshold branching ratios in the strangeness 
$S=-1$ channels~\cite{Oset:1998it,Oset:2001cn,Oller:2000fj}, where 
$a(630\text{ MeV}) \simeq -2$ is obtained. This is quite consistent with the
subtraction constant obtained by our procedure with the averaged mass of the
ground state octet baryons $M_T=1151$ MeV:
\begin{equation}
    a(630\text{ MeV}) \sim -1.98 \ ,
    \nonumber
\end{equation}
where we have used $a(\mu^{\prime})=a(\mu)
+2\ln(\mu^{\prime}/\mu)$~\cite{Hyodo:2003qa}.
In this way, we consider that the present renormalization scheme provides 
a natural extension to the channels in which the scattering observables are 
not available.

\subsection{Dynamically generated states and SU(3) limit}
\label{subsec:exmple}

When a baryon resonance $R$ is dynamically generated, the scattering 
amplitude has a pole in the second Riemann sheet of the complex energy 
plane. We can extract information of the resonance from the pole, by 
identifying the scattering amplitude close to the resonance energy region 
by the Breit-Wigner amplitude plus nonresonant background term:
\begin{align}
    -it_{ij}(\sqrt{s})
    &=
    -ig_i\frac{i}{\sqrt{s}-M_R+i\Gamma_R/2}(-ig_j)-it_{ij}^{BG} ,
    \nonumber
\end{align}
for an $s$-wave resonance in the coupled-channel scattering with mass $M_R$ 
and total width $\Gamma_R$, where $g_i$ is the coupling strength of the 
resonance $R$ to the channel $i$. Since the NG boson is scattered in an 
$s$-wave, dynamically generated states have the same spin but opposite 
parity with the target hadron.

In the literature, many experimentally observed resonances were identified 
as hadronic molecule states generated dynamically in the scattering of the 
NG boson and hadrons. The studies with the octet baryon target successfully 
reproduced the $J^P=1/2^-$ resonances $\Lambda(1405)$, $\Lambda(1670)$, 
$\Sigma(1620)$~\cite{Oset:2001cn}, $N(1535)$~\cite{Inoue:2001ip}, and 
$\Xi(1620)$~\cite{Ramos:2002xh}. In a different 
scheme~\cite{Garcia-Recio:2003ks}, the $\Xi(1690)$ resonance was found in 
addition to the $\Xi(1620)$. In the scattering with baryon decuplet 
target~\cite{Kolomeitsev:2003kt,Sarkar:2004jh}, the $J^P=3/2^-$ resonances 
such as $\Lambda(1520)$, $\Xi(1820)$, $\Sigma(1670)$ were well reproduced.
In the heavy sector, charmed resonances $\Lambda_c(2880)$ and 
$\Lambda_c(2593)$ were generated dynamically in the scattering of the NG 
boson with the ground state charmed baryons~\cite{Lutz:2003jw}. For the 
heavy meson sector, the recently discovered $D_s(2317)$ was properly 
reproduced as a resonance of the NG boson and the ground state charmed
mesons~\cite{Kolomeitsev:2003ac}.

It was shown that the resonances found in the light baryon sectors became 
bound states in the SU(3) limit, by tracing the positions of poles with 
gradual restoration of SU(3) symmetry~\cite{Jido:2003cb,Garcia-Recio:2003ks,
Sarkar:2004jh}. The existence of the bound state poles is also observed in 
the heavy sectors~\cite{Lutz:2003jw,Kolomeitsev:2003ac}. These facts imply 
that the bound states are first generated in the SU(3) limit; then they 
acquire widths through the coupled-channel dynamics in the different 
thresholds among the channels due to the SU(3) breaking. Therefore, in order
to clarify the mechanism to generate physical resonances, we study the bound
states in the SU(3) limit, which are expected to be the origin of the 
resonances observed in nature. 

In the next section, based on the group theoretical argument, we first study
the sign and coupling strength of the WT interaction systematically for 
various channels, which is the driving force to generate the bound state. Of
particular interest are the flavor exotic channels, where a small attraction
will be found. Then we discuss whether the bound states can be generated 
with the given target hadron mass and the coupling strength obtained in the 
group theoretical argument. 

\section{Weinberg-Tomozawa interaction}\label{sec:WTterm}

\subsection{Group theoretical structure}\label{susec:groupstructure}

\begin{figure}[b]
\centerline{\includegraphics[width=0.55\textwidth]{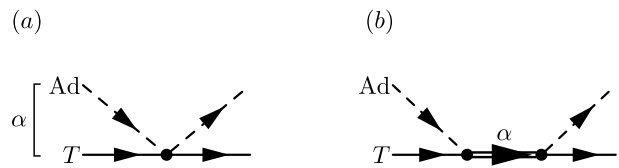}}
\caption{(a): Notation of the representations $\alpha$, Ad, and $T$ for 
the WT term. (b): The bound state pole diagram after unitarization of 
the amplitude.}
\label{fig:Rep}
\end{figure}

Here we discuss the interaction strengths $C_{\alpha ,T}$ in 
Eq.~\eqref{eq:WTint}. $C_{\alpha ,T}$ is the interaction strength of the WT 
term in the channel belonging to the SU(3) irreducible representation 
$\alpha$ in the direct product space of the target hadron $T$ and the NG 
boson Ad as shown in Fig.~\ref{fig:Rep}(a). The interaction strength is 
dependent only on the representations of the channel $\alpha$, the hadron 
$T$ and the NG boson Ad which is always fixed as the adjoint representation.
The WT term is the vector-current--vector-current interaction, so that 
$C_{\alpha,T}$ (with proper normalization) can be written as 
\begin{equation}
    C_{\alpha,T}
    =
    -2\bra{[\text{Ad},T]_{\alpha}}{\bm F}_{T} \cdot {\bm F}_{\rm Ad}
    \ket{[\text{Ad},T]_{\alpha}} ,
    \label{eq:WTstructure}
\end{equation}
where ${\bm F}_{T} $ and ${\bm F}_{\rm Ad}$ are the SU(3) generators in the 
representations of the hadron $T$ and the NG boson Ad, respectively. The 
state labelled by $[\text{Ad},T]_{\alpha}$ belongs to the representation
$\alpha$ composed of the two particle system $\text{Ad}\otimes T$, which is 
schematically shown in Fig.~\ref{fig:Rep}(a). The bound states appearing in
the channel $\alpha$ after the unitarization of the amplitude has the flavor
quantum number $\alpha$, as depicted in Fig.~\ref{fig:Rep}(b). 

The matrix element of Eq.~\eqref{eq:WTstructure} can be expressed in terms 
of Casimir invariants:
\begin{align*}
    &\bra{[\text{Ad},T]_{\alpha}}{\bm F}_{T} \cdot {\bm F}_{\rm Ad}
    \ket{[\text{Ad},T]_{\alpha}} \\
    &\quad =\tfrac{1}{2}\left[C_2(\alpha)-C_2(\text{Ad})-C_2(T)\right] ,
\end{align*}
where $C_2(R)$ is the quadratic Casimir of the representation $R$. This is 
consistent with the fact that the WT term is invariant under the SU(3) 
transformation. Recalling $C_{2}(\text{Ad}) = 3$ for the adjoint 
representation, we find that the strength of the WT interaction can be 
written as
\begin{equation}
    C_{\alpha,T}= C_2(T)-C_2(\alpha)+3 .
    \label{eq:WTintfinal}
\end{equation}
As seen in Eq.~\eqref{eq:WTint}, the potential is proportional to 
$-C_{\alpha,T}$, and therefore a negative $C_{\alpha,T}$ leads to a 
repulsive interaction, whereas a positive $C_{\alpha,T}$ gives an attractive
interaction in the present convention. The quadratic Casimir for the $[p,q]$
representation in the tensor notation is given by
\begin{equation}
    C_2([p,q])
    =\tfrac{1}{3}[p^2+q^2+pq+3(p+q)] .
    \label{eq:SU3Casimir}
\end{equation}
In the SU(2) case, Eq.~\eqref{eq:WTintfinal} is expressed as
\begin{equation}
    C_{\alpha,T}^{\text{SU(2)}}= -[I_{\alpha}(I_{\alpha}+1)-I_T(I_T+1)-2] ,
    \label{eq:WTintSU2}
\end{equation}
where $I_{\alpha}$ is the total isospin of the $\pi$-$T$ state, and $I_T$ 
is the isospin of the target. This is well-known expression of the 
$\pi N$ scattering lengths of the Weinberg-Tomozawa 
relation~\cite{Weinberg:1966kf,Tomozawa:1966jm}.

It is worth noting that, for the SU(3) case, the $C_{ij}$ in the isospin 
space are given by Eq.~\eqref{eq:WTintfinal} and the SU(3) isoscalar 
factors~\cite{PRSLA268.567,deSwart:1963gc} with suitable phase conventions:
\begin{align}
    & \hspace{-0.5cm} C_{ij}(I,I_{M_i}, Y_{M_i}, I_{T_i}, Y_{T_i},
    I_{M_j}, Y_{M_j}, I_{T_j}, Y_{T_j}) \nonumber \\
    =&\sum_{\alpha}[C_2(T)-C_2(\alpha)+3]
    \iso{8}{T}{\alpha}
    {I_{M_i} , Y_{M_i}}{I_{T_i} , Y_{T_i}}{I , Y} \nonumber \\
    &\times\iso{8}{T}{\alpha}
    {I_{M_j} , Y_{M_j}}{I_{T_j} , Y_{T_j}}{I , Y} ,
    \nonumber \\
     Y = &Y_{M_i} +Y_{T_i} = Y_{M_j} +Y_{T_j} ,
    \nonumber 
\end{align}
where $I_{X_i}$ and $Y_{X_i}$ are the isospin and hypercharge of the state 
$X$ in channel $i$. 

\subsection{General expression for coupling strength \\
and exoticness}

To form a bound state of a two-body system, the coupling should at least be 
attractive, $C_{\alpha,T}>0$, namely  $C_2(\alpha)<3+C_2(T)$. This indicates
the difficulty of generating exotic resonances from a nonexotic target, 
since the exotic states are generally in higher dimensional representations,
and their Casimir invariants have larger values than those of the simple 
nonexotic hadrons. 

Here we evaluate the coupling strengths $C_{\alpha,T}$, using 
Eqs.~\eqref{eq:WTintfinal} and \eqref{eq:SU3Casimir}. We first consider the 
target hadrons in specific representations, such as octet and decuplet. We 
then consider the general cases with arbitrary representations of the 
target hadrons.

Let us consider the light flavor baryons, where the lowest lying hadrons are
in octet ($N$, $\Lambda$, $\Sigma$, $\Xi$) and decuplet ($\Delta$, 
$\Sigma^{*}$, $\Xi^{*}$, $\Omega$) representations. The representations for 
$\alpha$ are found in the irreducible decomposition $T\otimes \bm{8} = \sum 
\alpha$. The coupling strengths for the octet $T=\bm{8}$ and decuplet 
$T=\bm{10}$ targets are obtained by~\eqref{eq:WTintfinal} and 
\eqref{eq:SU3Casimir} and summarized in Table~\ref{tbl:lightbaryon}; they 
were also reported in Refs.~\cite{Jido:2003cb,Kolomeitsev:2003kt,
Sarkar:2004jh}. For light flavor baryons, nonexotic representations are the 
singlet, octet and decuplet representations, since they can be achieved by 
the three light quarks $qqq$: $\bm{3}\otimes \bm{3} \otimes \bm{3} = 
\bm{1}\oplus \bm{8} \oplus \bm{8} \oplus \bm{10}$. Thus, for the light 
flavor baryons, $\alpha=\overline{\bm{10}}$, $\bm{27}$, and $\bm{35}$ are 
the exotic channels. We find that, among exotic channels, only the channel 
$\bm{27}$ composed by the decuplet target is attractive, 
$C_{\bm{27},\bm{10}}=1>0$~\cite{Sarkar:2004sc}, while the couplings in other
exotic channels are either repulsive or zero.

\begin{table}[tbp]
    \centering
    \caption{The coupling strengths $C_{\alpha,T}$ for light flavor 
    baryons.}
    \begin{ruledtabular}
    \begin{tabular}{ccccccc}
        $\alpha$ & $\bm{1}$ & $\bm{8}$ & $\bm{10}$ & $\overline{\bm{10}}$
	& $\bm{27}$ & $\bm{35}$  \\
        \hline
        $T=\bm{8}$ & 6 & 3 & 0 & 0 & $-2$ &   \\
        $T=\bm{10}$ &  & 6 & 3 &  & 1 & $-3$  \\
    \end{tabular}
    \end{ruledtabular}
    \label{tbl:lightbaryon}
\end{table}
\begin{table}[b]
    \centering
    \caption{The coupling strengths $C_{\alpha,T}$ for heavy flavor 
    hadrons.}
    \begin{ruledtabular}
    \begin{tabular}{cccccccc}
        $\alpha$ & $\overline{\bm{3}}$ & $\bm{6}$ & $\overline{\bm{15}}$ 
	& $\bm{24}$
	& $\bm{3}$ & $\overline{\bm{6}}$ & $\bm{15}$  \\
        \hline
        $T=\overline{\bm{3}}$ & 3 & 1 & $-1$ &  &  & &  \\
        $T=\bm{6}$ & 5 & 3 & 1 & $-2$ & & & \\
        $T=\bm{3}$ &  &  &  &  & 3 & 1 & $-1$  \\
    \end{tabular}
    \end{ruledtabular}
    \label{tbl:heavyhadron}
\end{table}

We can apply the same argument to hadrons with heavy quarks. In the SU(3) 
group argument, the heavy quarks are spectators. For the hadrons containing 
one heavy quark $Q$, the nonexotic baryons $qqQ$ are classified in 
$\overline{\bm{3}}$ and $\bm{6}$, and the mesons $q\bar{Q}$ are in $\bm{3}$.
The representation $\bm 3$ is also used for baryons with two heavy quarks, 
which are not experimentally well established to construct a multiplet. The 
established lowest lying hadrons are classified in these nonexotic 
multiplets: charmed baryons $\overline{\bm{3}}$ ($\Lambda_c$, $\Xi_c$) and 
$\bm{6}$ ($\Sigma_c$, $\Xi_c^*$, $\Omega_c$), charmed mesons $\bm 3$ ($D$, 
$D_s$), and bottomed mesons $\bm 3$ ($B$, $B_s$). With these states being 
target hadrons, coupling strengths $C_{\alpha,T}$ are evaluated as shown in 
Table~\ref{tbl:heavyhadron}. For heavy baryons, $\overline{\bm{15}}$ and 
$\bm{24}$ are the exotic channels, while $\overline{\bm{6}}$ and $\bm{15}$ 
are the exotic states for the heavy mesons. We find that, as seen in 
Table~\ref{tbl:heavyhadron}, the interaction in the exotic channel $\bm{15}$
composed of the $\bm{6}$ heavy baryon is attractive, $C_{\overline{\bm{15}},
\bm{6}}=1>0$~\cite{Lutz:2003jw}, and the mesonic channel $\bm 6$ from the 
$\bm 3$ heavy meson target is also attractive, $C_{\overline{\bm{6}},\bm{3}}
=1>0$~\cite{Kolomeitsev:2003ac}. Note that the states in 
$\overline{\bm{15}}$ and $\bm{24}$ are flavor exotic with one heavy quark, 
which are different from the $\Theta_c$ state of charm $-1$ discussed in 
Refs.~\cite{Aktas:2004qf,Cohen:2005bx,Lee:2005pn}, since it has one 
anticharm quark.

It is interesting that the attractive interactions for the exotic 
channels are found only in the limited cases $(\alpha ,T)=
(\bm{27},\bm{10})$, $(\overline{\bm{15}},\bm{6})$, and 
$(\overline{\bm{6}},\bm{3})$ with a universal strength $C_{\alpha,T}=1$. 
In the following, we show that this observation is true for the 
general targets with arbitrary representations.

\begin{table*}[bt]
    \centering
    \caption{Properties of the WT interaction for various channels of 
    representations $\alpha$ formed by a target hadron $[p,q]$ and an octet 
    meson $[1,1]$. Tabulated from the left to right column are the channel
    representation $\alpha$, the condition to have $\alpha$ in the 
    irreducible decomposition, the coupling strengths of the WT term 
    $C_{\alpha,T}$ and its sign, the differences of the exoticness $E$, 
    $\epsilon$, $\nu$ between $\alpha$ and $T$, the coupling strengths with 
    arbitrary $N_c$, $C_{\alpha,T}(N_c)$, and the sign of the WT interaction
    at the large-$N_c$ limit.}
    \begin{ruledtabular}
    \begin{tabular}{ccccccccc}
	$\alpha$ & Condition 
	& $C_{\alpha,T}$ & Sign 
	& $\Delta E$ & $\Delta \epsilon$ & $\Delta \nu$ 
	& $C_{\alpha,T}(N_c)$ & $V(N_c\to \infty)$   \\
        \hline
        $[p+1,q+1]$ & -
	& $-p-q$ & Repulsive 
	& 1 or 0 & 1 & 0
	& $\frac{3-N_c}{2}-p-q$ & Repulsive \\
        $[p+2,q-1]$ & $q\geq 1$ 
	& $1-p$ & \dots
	& 1 or 0 & 0 & 1 
	& $1-p$ & $0$ \\
        $[p-1,q+2]$ & $p\geq 1$ 
	& $1-q$ & \dots 
	& 1 or 0 & 1 & $-1$ 
	& $\frac{5-N_c}{2}-q$ & Repulsive \\
        $[p,q]$ & $q\geq 1$ 
	& $3$ & Attractive 
	& 0 & 0 & 0 
	& 3 & $0$ \\
        $[p,q]$ &  $p\geq 1$  
	& $3$ & Attractive 
	& 0 & 0 & 0 
	& 3 & $0$\\
        $[p+1,q-2]$ & $q\geq 2$
	& $3+q$ & Attractive
	& 0 or $-1$ & $-1$ & $1$ 
	& $\frac{3+N_c}{2}+q$ &  Attractive \\
        $[p-2,q+1]$ & $p\geq 2$ 
	& $3+p$ & Attractive 
	& 0 or $-1$ & 0 & $-1$ 
	& $3+p$ & $0$ \\
        $[p-1,q-1]$ & $p\geq 1$, $q\geq 1$ 
	& $4+p+q$ & Attractive 
	& 0 or $-1$ & $-1$ & $0$
	& $\frac{5+N_c}{2}+p+q$ & Attractive\\
    \end{tabular}
    \end{ruledtabular}
    \label{tbl:ECtable}
\end{table*}

Let us consider an arbitrary representation $T=[p,q]$ for the target hadron.
Possible representations $\alpha$ for the hadron and NG boson system are 
obtained in the irreducible decomposition of the direct product of the 
$[p,q]$ for the target hadron and  the adjoint $[1,1]$ for the NG boson:
\begin{align}
    [p,q]\otimes [1,1] 
    =&    [p+1,q+1]   
    \oplus [p+2,q-1]\nonumber \\  
    &\oplus [p-1,q+2]
    \oplus [p,q] 
    \oplus [p,q]  \nonumber \\
    &\oplus [p+1,q-2]
    \oplus [p-2,q+1] \nonumber \\
    &\oplus [p-1,q-1] .
    \label{eq:generalpq}
\end{align}
There are  maximally eight representations. It is understood that the 
representation $[a,b]$ with $a< 0$ or $b< 0$ is not taken into account, and 
that, for two $[p,q]$ representations, one has the condition $p\geq 1$ and 
the other has $q\geq 1$. The coupling strengths $C_{\alpha,T}$ are 
summarized in the third column of Table~\ref{tbl:ECtable}. This is the 
general expression for the coupling strength of the WT interaction for 
arbitrary representations. As seen in the table, since $p$ and $q$ are 
nonnegative, the sign of the interaction is determined for a given $\alpha$,
except for $[p+2,q-1]$ and $[p-1,q+2]$. The interactions of these channels 
are either attractive or repulsive, depending on the values of $p$ and $q$.
It is worth noting that $C_{\alpha,T}$ is an integer, since $p$ and $q$ are 
also integers.

In the general expression \eqref{eq:generalpq}, it is not known which
representations are exotic before specifying the baryon number of the target
hadron. In order to discuss the exotic states, we define the exoticness 
$E$~\cite{Kopeliovich:1990ez,Diakonov:2003ei,Kopeliovich:2003he,
Jenkins:2004tm} as the number of valence quark-antiquark pairs to compose 
the flavor multiplet. We denote $B$ as the baryon number carried by $u$, 
$d$, and $s$ quarks. The number of the heavy quarks is not counted as the 
baryon number here. For $B>0$, the exoticness $E$ is given by
\begin{align}
    E=\epsilon\theta(\epsilon)+\nu\theta(\nu) ,
    \label{eq:exoticness} 
\end{align}
where 
\begin{equation}
    \epsilon \equiv
    \dfrac{p+2q}{3}-B, \quad 
    \nu\equiv\dfrac{p-q}{3}-B .
    \nonumber
\end{equation}
Note that $\epsilon \geq \nu$. More general arguments and derivations are 
given in Appendix~\ref{sec:Exoticness}.

We are interested in the channel $\alpha$ which has a larger exoticness $E$ 
than that of the target hadron $T$. To find it we define 
\begin{equation}
    \Delta \epsilon =\epsilon_{\alpha}- \epsilon_{T} ,
    \quad \Delta \nu=\nu_{\alpha}- \nu_{T} ,
    \nonumber
\end{equation}
which are shown in the fifth and sixth columns of Table~\ref{tbl:ECtable}.
With these quantities, we evaluate the difference of the exoticness 
$\Delta E=E_{\alpha}- E_{T}$:
\begin{equation}
    \Delta E=
    \begin{cases}
	0
	& \text{for}\quad  \epsilon_\alpha \leq 0 , \ \epsilon_T \leq 0  \\
	\Delta \epsilon
	& \text{for}\quad  \nu_\alpha \leq 0 , \ \nu_T \leq 0  \\
	\Delta\epsilon +\Delta\nu   & \text{for}\quad \text{others}
	\end{cases} .
    \nonumber
\end{equation}
Therefore, the condition to generate the states with larger exoticness than 
the target ($\Delta E=1$) is given by one of the following cases:
\begin{enumerate}
    \item[(i)]  $\Delta \epsilon=1$, $\Delta \nu = 0$, $\epsilon_{T}\geq 0$,
    \item[(ii)]  $\Delta \epsilon=0$, $\Delta \nu = 1$,
    $\nu_{T}\geq 0$,
    \item[(iii)]  $\Delta \epsilon=1$, $\Delta \nu = -1$,
    $\nu_{T}\leq 0$.
\end{enumerate}
These cases correspond to the first three rows of Table~\ref{tbl:ECtable}.
For the target in $[p,q]$, case (i) is satisfied for $\alpha=[p+1,q+1]$, 
but the interaction is always repulsive for this channel.  Case (ii) is 
satisfied for $\alpha=[p+2,q-1]$. In order to have attraction, however, 
$p=0$ is required, which leads to $B\leq -q/3$ because of $\nu_{T}\geq 0$. 
This is achieved by the hadrons with negative baryon 
number, which are not considered here. Case (iii) is satisfied for 
$\alpha=[p-1,q+2]$, where the interaction can be attractive only when $q=0$ 
and the strength is $C_{\alpha,T}=1$. In this case, the condition 
$\nu_T\leq 0$ gives $p\geq 3B$.

What we have shown here is that the attractive interaction in more 
``exotic'' channel than the target hadron is only found as
\begin{equation}
    C_{\text{exotic}}=1 ,
    \label{eq:Cexotic}
\end{equation}
with
\begin{equation}
    T=[p,0], \quad
    \alpha=[p-1,2] , \quad
    p\geq 3B .
    \nonumber
\end{equation}
It is interesting that the strength is always $C_{\alpha,T}=1$, which is the
smallest strength of the Weinberg-Tomozawa term. In addition, this takes 
place when the target hadron belongs to the totally symmetric representation
$[p,0]$. The examples shown above for the ground states are the special 
cases of this conclusion with $p=1$, 2, and 3 ($T=\bm{3}$, $\bm{6}$, and 
$\bm{10}$, respectively). 

Let us consider the exoticness of the representation $T=[p,0]$ with
$p\geq 3B$,
\begin{equation}
    E=\left(\frac{p}{3}-B\right)+\left(\frac{p}{3}-B\right)
    =2\left(\frac{p}{3}-B\right) .
    \nonumber
\end{equation}
It follows from the condition of triality~\eqref{eq:triality} that $(p/3-B)$
is an integer. This means that the attractive channels to generate more 
exotic states appear for even number of $E$. For instance, for $B=1$ 
hadrons, $[3,0]$ is $E=0$ and $[6,0]$ is $E=2$, and there is no such channel
in $E=1$. Thus, even if the attraction~\eqref{eq:Cexotic} is enough to 
generate a bound state, the WT interaction can generate at most 
$\Delta E=+1$ state, and it is not possible to generate a tower of exotic 
states recursively. 

The next question is whether the attraction~\eqref{eq:Cexotic} is 
strong enough to provide a bound state. This will be discussed in 
Sec.~\ref{sec:Boundstate}.

\subsection{Large-$N_c$ limit of the WT term}

In this section we discuss the large-$N_c$ limit of the WT term. It is known
that the WT term scales as $\mathcal{O}(1/N_c)$ in the large-$N_c$ limit, 
since it contains $1/f^2$ and $f\sim
\mathcal{O}(\sqrt{N_c})$~\cite{Hooft:1974jz,Witten:1979kh}. However, if the 
coupling strength $C_{\alpha,T}$ has some $N_c$ dependence, the scaling
of the amplitude will be different. In Ref.~\cite{Garcia-Recio:2006wb}, 
nontrivial $N_c$ dependence of the WT term was reported for the spin-flavor 
SU(6) extended WT term. Here we show that $C_{\alpha,T}$ does have the $N_c$
dependence in the case of the baryon target without incorporating the spin 
degrees of freedom. The representations of mesons do not depend on $N_c$ so 
that the result for the heavy meson target remains unchanged. An interesting
discussion was recently made on interplay of the chiral and $1/N_c$ 
expansions~\cite{Cohen:2006up}. In the present argument, we perform the 
chiral expansion first, then consider the large-$N_c$ expansion.

For arbitrary $N_c$, a baryon is constructed by $N_c$ quarks and $E$ pairs 
of quarks and antiquarks, where $E$ is the exoticness. Accordingly, the 
SU(3) flavor representations for the baryon are extended
as~\cite{Dashen:1993jt,Cohen:2004ki}
\begin{equation}
    [p,q] \to \left[p,q+\frac{N_c-3}{2}\right] ,
    \label{eq:largeNrep}
\end{equation}
which reduces to $[p,q]$ at $N_c=3$. In this extension, the baryon spin is 
implicitly fixed at the value of $N_c=3$. Denoting the $N_c$ extended 
representation of $\bm{R}$ as ``$\bm{R}$'', we find that the coupling 
strengths $C_{\largeN{\alpha},\largeN{T}}$ with arbitrary $N_c$ can be given
by
\begin{equation}
    C_{\largeN{\alpha},\largeN{T}} 
    =C_{2}(\largeN{T})-C_{2}(\largeN{\alpha})+3 .
    \label{eq:WTintNc}
\end{equation}
Note that the representation of the meson does not change in the large-$N_c$
limit. The general form of the quadratic Casimir is given in 
Eq.\eqref{eq:SU3Casimir}. The Casimir of the large-$N_{c}$ baryon 
``$[p,q]$'' is given by
\begin{align}
    C_2(\largeN{[p,q]})
    =&C_2\left(\left[p,q+\frac{N_c-3}{2}\right]\right) \nonumber \\
    =&\frac{1}{3}\left(
    -\frac{9}{4}+p^2+\frac{3p}{2}+pq+q^2\right)\nonumber \\
    &+\frac{1}{3}\left(\frac{p}{2}+
    q\right)
    N_c+\frac{N_c^2}{12} ,
    \label{eq:largeNcCashimir}
\end{align}
for arbitrary $N_c$. Note that $C_2(\largeN{[p,q]})\neq 
C_2(\largeN{[q,p]})$, because $\largeN{[p,q]}\neq \overline{\largeN{[q,p]}}$
for $N_c\neq 3$ by the construction of the large-$N_{c}$ 
baryon~\eqref{eq:largeNrep}. It is important to note also that the 
coefficient of the linear term of $N_{c}$ in the right-hand side of 
Eq.~\eqref{eq:largeNcCashimir} depends on the representation $[p,q]$, while 
the coefficient of $N_c^2$ is independent of the representation. This means 
that there is nontrivial linear $N_{c}$ dependence at order 
$\mathcal{O}(N_c)$ in the coefficient of the WT term: $C_{\largeN{\alpha},
\largeN{T}}(N_c)=C_2(\largeN{T})-C_2(\largeN{\alpha})+3$, where the leading 
$N_c^2$ terms are canceled. This is consistent with the general $N_c$ 
dependence $\mathcal{O}(N_c^0)$ of the meson-baryon interaction.

For specific target hadrons, using Eqs.~\eqref{eq:WTintNc} and 
\eqref{eq:largeNcCashimir} we evaluate the $C_{\largeN{\alpha},\largeN{T}}
(N_c)$ coefficients with arbitrary $N_c$. These are summarized in 
Table~\ref{tbl:lightbaryonN} for light flavor baryons and in 
Table~\ref{tbl:heavyhadronN} for heavy flavor hadrons. It is worth noting 
that the attraction of $C_{\largeN{\bm{27}},\largeN{\bm{10}}}$ with 
$N_{c}=3$ turns into repulsion for $N_c>5$, which is also the case for 
$C_{\largeN{\overline{\bm{15}}},\largeN{\bm{6}}}$. Thus, the attractions 
found in exotic baryon channels with $N_{c}=3$ change into repulsive 
interaction at large $N_c$. Moreover, the repulsion in the exotic channels 
linearly increases in the large-$N_{c}$ limit. It is also interesting that 
the attraction in the coupling  $C_{\largeN{\bm{1}},\largeN{\bm{8}}}$ 
linearly increases as $N_c$ is increased. As mentioned above, the heavy 
meson $q\bar{Q}$ does not change its representation in the large-$N_c$ 
limit. Therefore the strengths shown in Table~\ref{tbl:heavyhadron} hold in 
arbitrary $N_c$, and the  exotic channel $\alpha =\overline{\bm{6}}$ 
remains attractive, as shown in Table~\ref{tbl:heavyhadronN}.

\begin{table}[bp]
    \centering
    \caption{The coupling strengths $C_{\largeN{\alpha},\largeN{T}}(N_c)$
    for light flavor baryons.}
    \begin{ruledtabular}
    \begin{tabular}{ccccccc}
        $\alpha$ & $\largeN{\bm{1}}$ & $\largeN{\bm{8}}$ 
	& $\largeN{\bm{10}}$ & $\largeN{\overline{\bm{10}}}$
	& $\largeN{\bm{27}}$ & $\largeN{\bm{35}}$  \\
        \hline
        $T=\largeN{\bm{8}}$ & $\frac{9}{2}+\frac{N_c}{2}$ & 3 & 0 
	& $\frac{3}{2}-\frac{N_c}{2}$ & $-\frac{1}{2}-\frac{N_c}{2}$ &   \\
        $T=\largeN{\bm{10}}$ &  & 6 & 3 &  & $\frac{5}{2}-\frac{N_c}{2}$
	& $-\frac{3}{2}-\frac{N_c}{2}$  \\
    \end{tabular}
    \end{ruledtabular}
    \label{tbl:lightbaryonN}
    \caption{The coupling strengths $C_{\largeN{\alpha},\largeN{T}}(N_c)$
    for heavy flavor hadrons. $T=\bm{3}$ is assigned to heavy mesons so 
    that $C_{\alpha,T}(N_c)$ are unchanged.}
    \begin{ruledtabular}
    \begin{tabular}{cccccccc}
	$\alpha$ & $\largeN{\overline{\bm{3}}}$ & $\largeN{\bm{6}}$
	& $\largeN{\overline{\bm{15}}}$ & $\largeN{\bm{24}}$
	& $\bm{3}$ & $\overline{\bm{6}}$ & $\bm{15}$  \\
        \hline
        $T=\largeN{\overline{\bm{3}}}$ & 3 & 1 
	& $\frac{1}{2}-\frac{N_c}{2}$ &&&& \\
        $T=\largeN{\bm{6}}$ & 5 & 3 & $\frac{5}{2}-\frac{N_c}{2}$ 
	& $-\frac{1}{2}-\frac{N_c}{2}$ & & & \\
        $T=\bm{3}$ &  &  &  &  & 3 & 1 & $-1$  \\
    \end{tabular}
    \end{ruledtabular}
    \label{tbl:heavyhadronN}
\end{table}

The above discussion is on the large-$N_c$ behavior of the coupling
strengths. Recalling that the WT term has $N_c$ dependence of $1/f^2$, the
scaling of the amplitude of the WT term is in total given by the product of 
$1/N_c$ and $C_{\largeN{\alpha},\largeN{T}}(N_c)$. Therefore, in 
large-$N_c$ limit, the WT interaction
\begin{enumerate}
    \item[(i)] is attractive for the $\largeN{\bm{1}}$ channel in the 
    scattering of the $\largeN{\bm{8}}$ baryon and the NG boson,
   
    \item[(ii)] has no interaction for other nonexotic baryonic channels,

    \item[(iii)] is repulsive for all exotic baryonic channels,

    \item[(iv)] has no interaction for nonexotic and exotic mesonic 
    channels.
\end{enumerate}

The first statement is very interesting in the context of the two-pole 
structure of $\Lambda(1405)$~\cite{Jido:2003cb,Hyodo:2003jw,Magas:2005vu}.
It was found that two poles originate in the singlet and octet bound states 
in the SU(3) limit~\cite{Jido:2003cb}. Although the SU(3) breaking causes 
admixture of the singlet and octet states, different behavior of the poles
is expected in the large-$N_c$ limit. The investigation of the large-$N_c$ 
limit would provide deeper understanding of the $\Lambda(1405)$ resonance. 

We show that the coupling in the $\largeN{\bm{1}}$ channel is attractive in
the large-$N_c$ limit, while $\largeN{\overline{\bm{10}}}$, 
$\largeN{\bm{27}}$, and $\largeN{\bm{35}}$ channels are all repulsive. This
observation may be related to the kaon bound state approach to the 
Skyrmion~\cite{Callan:1985hy} which is essentially based on the large-$N_c$
limit. In this approach, $s$-wave bound states of $K$ with the SU(2) soliton
are obtained for $S=-1$ channel, which corresponds to $\Lambda(1405)$ and 
may largely have the flavor singlet component. On the other hand, exotic 
$S=+1$ states are not bound~\cite{Itzhaki:2003nr,Walliser:2005pi}. The 
relation between two different approaches in the large-$N_c$ limit is worth 
investigating, although the present analysis based on the SU(3) symmetric 
limit, while the bound state approach is based on the large SU(3) breaking.

We can also extend the large-$N_c$ limit to the general target hadrons 
belonging to the representation $[p,q]$. In the seventh column of 
Table~\ref{tbl:ECtable}, we show the coupling strength at arbitrary $N_c$. 
Combining with the $1/N_c$ factor of $1/f^2$, the sign of the WT interaction
at the large-$N_c$ limit can be obtained in the eighth column of 
Table~\ref{tbl:ECtable}. It is seen that in the large-$N_c$ limit, among the
exotic channels, $\alpha=[p+1,q+1]$ and $\alpha=[p-1,q+2]$ are repulsive and
the strength of the interaction for $\alpha=[p-1,q+2]$ is zero. In this way,
we show that no attractive interaction exists for exotic channels in the 
large-$N_c$ limit. This means that, in the large-$N_c$ limit, the 
nonexistence of the $s$-wave exotic
baryons can be shown without solving the scattering problem.

\section{Bound state solutions}\label{sec:Boundstate}

\subsection{Condition to generate a bound state}\label{subsec:Bound}

We have been discussing the kernel interaction of the chiral unitary 
approach so far and have found the possible attractive interaction in exotic
channels with universal strength $C_{\text{exotic}}=1$. Now we study the 
unitarized amplitude~\eqref{eq:ChUamp} with the WT interaction, and derive 
the condition to generate a bound state pole. In other words, we solve the 
Schr\"odinger equation with the potential~\eqref{eq:WTint}. Before studying 
the chiral unitary approach, let us note that the WT interaction is 
independent of the three-momentum, as seen in Eq.~\eqref{eq:WTint}. 
Therefore it is instructive to recall an analogous problem of the delta 
function potential in three-dimensional nonrelativistic quantum mechanics.

As is well known, in one spatial dimension, the delta function potential
always provides one bound state, if the interaction is attractive. However,
in three dimensions, the existence of a bound state is not trivial. 
Moreover, an ultraviolet divergence appears in obtaining the wave function 
in coordinate space, as we show in Appendix~\ref{sec:Delta}. This is because
the eigenvalue problem is ill defined, since the short distance behavior of 
the potential is more singular than the kinetic term~\cite{Jackiw:1991je}. 
To obtain a physically meaningful result, we should tame the divergence by a
proper regularization scheme. With a three-momentum cutoff $\Lambda$, the 
binding energy of a bound state $E_b$ in the three-dimensional delta 
function potential $V(\bm{x})=-v\delta(\bm{x})$ is given by the equation
\begin{equation}
    \frac{1}{2mv}=\frac{1}{2\pi^2}
    \left(
    \Lambda -\sqrt{2mE_b}\arctan \left[\frac{\Lambda}{\sqrt{2mE_b}}\right]
    \right) ,
    \nonumber
\end{equation}
where $m$ is the reduced mass of the system (see Appendix~\ref{sec:Delta}).
Note that the binding energy depends on the cutoff $\Lambda$, and the 
solution does not always exist. The condition to have a bound state solution
is
\begin{align}
    v>\frac{\pi^2}{m\Lambda} \equiv v_c ,
    \nonumber
\end{align}
where we have defined the critical strength of attraction $v_c$. For a given
cutoff $\Lambda$, $v_c$ is the smallest attraction that can provide the 
bound state. If the attraction is less than $v_c$, no bound state exists. In
addition, as shown in Appendix~\ref{sec:Delta}, no resonance solution with a
complex energy is found in the delta function potential.

Turning to the chiral unitary approach, the problem is quite similar to the 
delta function potential problem, but with the following differences: (i) 
relativistic kinematics and (ii) energy dependence of the coupling. In order
to find the bound state in the scattering amplitude~\eqref{eq:ChUamp}, we 
write the denominator of the amplitude as
\begin{equation}
    D(\sqrt{s})
    \equiv 1-V(\sqrt{s})G(\sqrt{s}) .
    \label{eq:Ddef}
\end{equation}
Then the mass of the bound state with its mass $M_b$ is obtained by
\begin{equation}
    D(M_b)=0 
    ,\quad M_T< M_b < M_T+m .
    \label{eq:boundsolution}
\end{equation}
The mass of the bound state should be between the target mass $M_T$ and the 
scattering threshold $m+M_T$. Let us consider the behavior of the 
$D(\sqrt{s})$ in this region. It follows from Eq.~\eqref{eq:WTint} that 
$V(M_T)=0$, and in the present renormalization 
condition~\eqref{eq:regucond}, $G$ also vanishes at $\sqrt{s}=M_T$, so that 
\begin{equation}
    D(M_T)=1 .
    \nonumber
\end{equation}
From Eqs.~\eqref{eq:WTint} and \eqref{eq:loop_s}, it is seen that, for the
attractive interaction $C_{\alpha,T}>0$, both $V(\sqrt{s})$ and 
$G(\sqrt{s})$ are monotonically decreasing in the region 
$M_T<\sqrt{s}<M_T+m$. Since $V(M_T)=G(M_T)=0$, $VG$ is positive and 
monotonically increasing for $M_T< \sqrt{s} < M_T+m$, and therefore we find
that $D=1-VG$ is monotonically decreasing in this region. This means that 
there is only one bound state, if it exists, and the condition to satisfy 
Eq.~\eqref{eq:boundsolution} is given by $D(M_T+m)<0$. Thus, we define the 
critical strength of attraction $C_{\text{crit}}$ such that
\begin{equation}
    D(M_T+m)=0 .
    \label{eq:WTcriticalcond}
\end{equation}
If the coupling strength is smaller than $C_{\text{crit}}$, no bound state 
exists. Substituting Eq.~\eqref{eq:WTint} into 
Eq.~\eqref{eq:WTcriticalcond}, we obtain
\begin{align}
    C_{\text{crit}}= \frac{2f^2 }{m[-G(M_{T}+m)]} ,
    \label{eq:WTcritical}
\end{align}
where
\begin{align}
    G(M_T+m)
    =&   
    \frac{2M_T}{(4\pi)^{2}}
    \left[
    a(M_T)+\frac{m}{m+M_T}
    \ln\frac{m^{2}}{M_T^{2}} 
    \right] ,
    \nonumber
\end{align}
with $\bar q=0$ at the threshold and $a(M_T)$ defined in 
Eq.~\eqref{eq:adet}. Note that $C_{\text{crit}}$ is a function of the mass 
of the target hadron $M_T$ once the mass $m$ and the decay constant $f$ of 
the NG boson are fixed.

It is worth noting that $V$ changes the sign at $\sqrt{s}=M_T$. The 
attractive WT interaction turns into a repulsive one for $\sqrt{s}<M_T$. 
This energy region corresponds to the kinematics of the (bound region of) 
$u$-channel scattering, which is not considered in the present formulation, 
since it is an unphysical region of the $s$-channel scattering. To study the
amplitude of this region properly, we should include the $u$-channel 
multiscattering diagrams in the scattering equation or introduce the effect
from the left-hand cut in the N/D method.

\subsection{Bound state spectrum}

Let us evaluate numerically the function $D(\sqrt{s})$ defined in 
Eq.~\eqref{eq:Ddef}, in order to find the energy of the bound state $M_b$ by
\begin{equation}
    D(M_b)=0  .
    \label{eq:binding}
\end{equation}
For the numerical computation, we use the decay constant $f=93$ MeV and the 
mass of the NG bosons $m=368$ MeV which is the averaged value over the 
pseudoscalar octet mesons. We first choose the target hadron masses $M_T$ 
by averaging over the masses of the experimentally known ground states of 
the light flavor baryons, the charmed baryons, and the $D$ and $B$ mesons 
given by Particle Data Group (PDG)~\cite{\PDG}. These are presented in the 
third column of Table~\ref{tbl:binding}. 

\begin{table}[tbp]
    \centering
    \caption{Masses $M_T$ and coupling constants $C_{\alpha,T}$ for several 
    targets in the SU(3) limit. The masses of bound states $M_b$ are 
    obtained by solving Eq.~\eqref{eq:binding} numerically.}
    \begin{ruledtabular}
    \begin{tabular}{lccrccc}
        Target hadron & $T$ & $M_T$ (MeV) & $\alpha$ & $C_{\alpha,T}$ 
	& $M_b$ (MeV)& $E_{b}$ (MeV)  \\
        \hline
        Light baryon & $\bm{8}$ & 1151 & $\bm{1}$ & $6$ & 1450 & 69  \\
        && & $\bm{8}$ & $3$ & 1513 & \phantom{0}7 \\
	& $\bm{10}$ & 1382 & $\bm{8}$ & $6$ & 1668 & 80 \\
        && & $\bm{10}$ & $3$ & 1737 & 13 \\
        && & $\bm{27}$ & $1$ & No solution & \\
        Charmed baryon & $\overline{\bm{3}}$ 
	& 2408 & $\overline{\bm{3}}$ & $3$ & 2736 & 40 \\
        && & $\bm{6}$ & $1$ & No solution  \\
	& $\bm{6}$ & 2534 & $\overline{\bm{3}}$ & $5$ & 2804 & 98 \\
        && & $\bm{6}$ & $3$ & 2860 & 42 \\
        &&  & $\overline{\bm{15}}$ & $1$ & No solution  \\
	$D$ meson & $\bm{3}_c$ & 1900 & $\bm{3}$ & $3$ & 2240 & 28 \\
        && & $\overline{\bm{6}}$ & $1$ & No solution  \\
	$B$ meson & $\bm{3}_b$ & 5309 & $\bm{3}$ & $3$ & 5600 & 77 \\
        && & $\overline{\bm{6}}$ & $1$ & No solution  \\
    \end{tabular}
    \end{ruledtabular}
    \label{tbl:binding}
\end{table}

\begin{figure*}[tbp]
\centerline{\includegraphics[width=\textwidth]{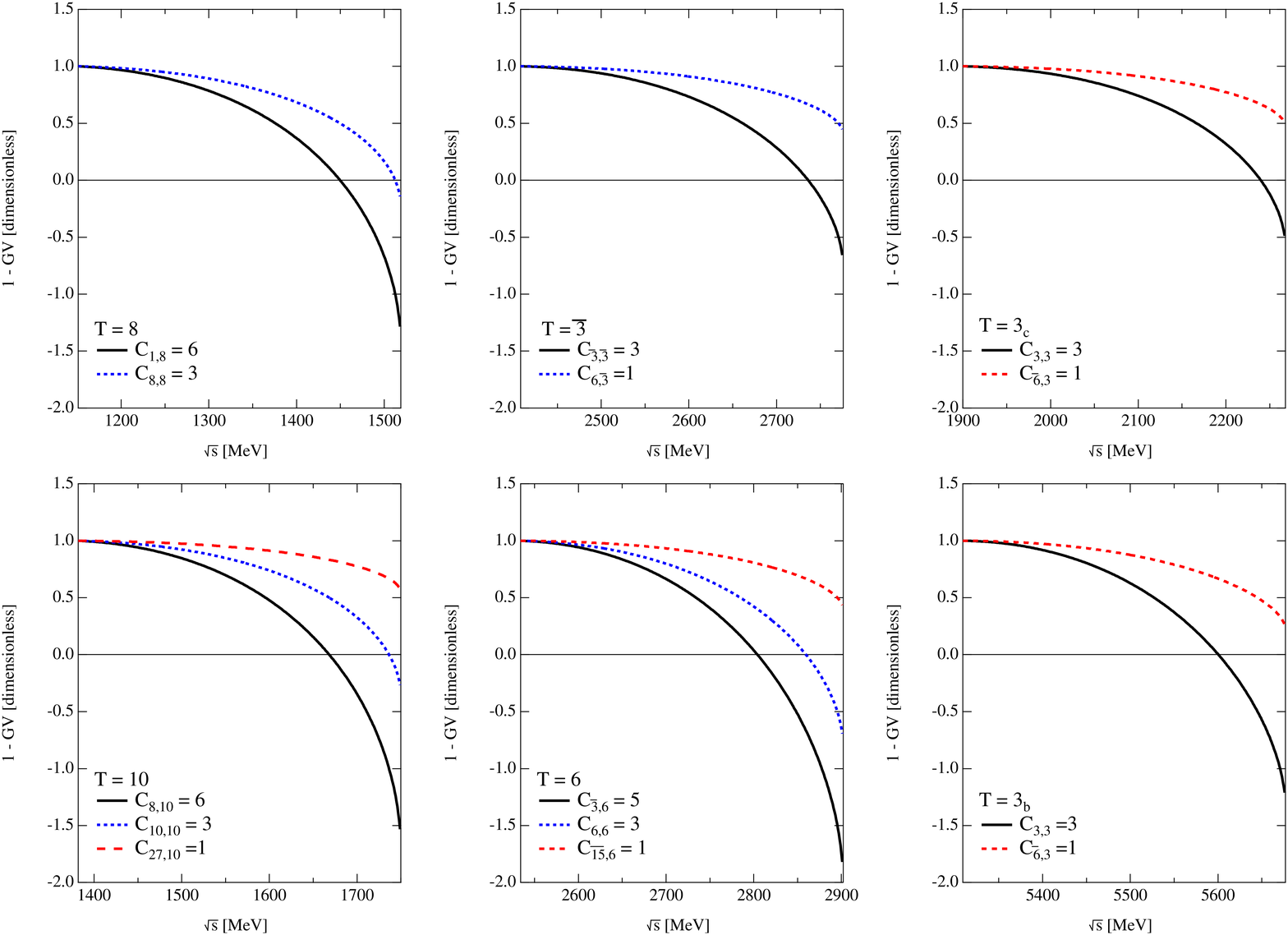}}
\caption{(color online). The denominator of the scattering amplitude of 
$1-VG$ for light flavor baryons (top left panel: $T=\bm{8}$, bottom left 
panel: $T=\bm{10}$), for charmed baryon (top center panel: 
$T=\overline{\bm{3}}$, bottom center panel: $T=\bm{6}$) and for $D$ and $B$ 
mesons (top right panel: $T=\bm{3}_c$, bottom right panel: $T=\bm{3}_b$). 
Results for exotic channels are denoted by 
dashed lines.}
\label{fig:1mVGplot}
\end{figure*}

In Fig.~\ref{fig:1mVGplot}, we plot $D=1-VG$ for various target hadrons with
the coupling strengths found in Sec.~\ref{susec:groupstructure}. The 
position where curves cross zero determine bound state energies. The 
resulting energies of bound states $M_b$ and the binding energies
\begin{equation}
    E_b=M_T+m-M_b
    \nonumber
\end{equation}
are summarized in the sixth and seventh column of Table~\ref{tbl:binding}. 
As expected, larger coupling strengths provide larger binding energies. 
Phenomenology of the bound states found here has been extensively studied 
in more realistic calculations with SU(3) breaking, and has been shown to 
reproduce experimentally observed resonances~\cite{Oset:2001cn,Inoue:2001ip,
Ramos:2002xh,Garcia-Recio:2003ks,Sarkar:2004jh,Lutz:2003jw,
Kolomeitsev:2003ac}. It is worth noting that no bound state is found for the
flavor exotic channels. The attraction in these channels $C_{\alpha,T}=1$ is
not enough to bind the two-body system for the physical masses of these 
hadrons. This point will be further studied in next subsection.

\begin{figure}[tbp]
\centerline{\includegraphics[width=0.6\textwidth]{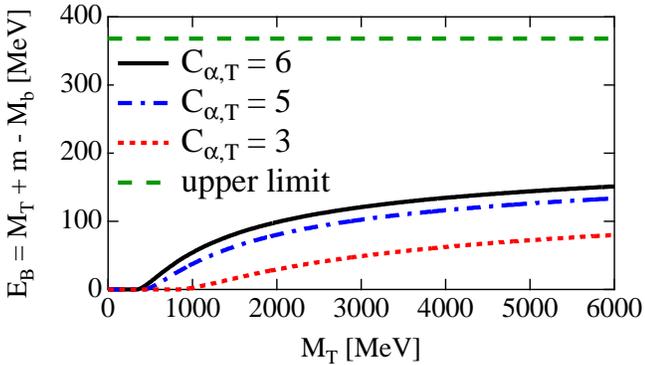}}
\caption{(color online). Binding energies $E_b=M_T+m-M_b$ as functions of 
$M_T$ with coupling strengths $C_{\alpha,T}=6$ (solid line), $5$ 
(dash-dotted line), and $3$ (dotted line). For reference, the upper limit of
the binding energies is indicated as the dashed line. No bound state is 
found for $C_{\alpha,T}=1$ in this energy region.} 
\label{fig:bindingenergy}
\end{figure}

It is instructive to show the dependence of the binding energy on the target
mass. In Fig.~\ref{fig:bindingenergy} we plot the binding energies 
$E_b=M_T+m-M_b$ as functions of the mass of target $M_T$, fixing the 
coupling strengths as $C_{\alpha,T}=6$, $5$, and $3$. $E_b=0$ corresponds to
the bound state exactly at the threshold, and $E_b=m$ corresponds to 
$M_b=M_T$. We find that the larger mass of the target provides larger 
binding energy. Note that $C_{\alpha,T}=1$ does not generate a bound state 
in this energy region $\sqrt{s}<6$ GeV.

The spectrum of the bound states in the light flavor sector shows an 
interesting structure, when we compare it with that of the SU(6) quark 
model. Recall the ground states and first excited states in the SU(6) quark 
model, which are $\bm{56}$ and $\bm{70}$, respectively. These 
representations include the spin-flavor quantum numbers of 
$\{ {}^{2S+1}D_{\text{SU(3)}}={}^2\bm{8}$, $^4\bm{10}\}$ in $\bm{56}$, and 
$\{ {}^2\bm{1}$, ${}^2\bm{8}$, $^2\bm{8}$, $^4\bm{1}$, $^4\bm{8}\}$ in 
$\bm{70}$. In the chiral unitary approach, the bound states of the baryon 
and NG boson have the same spin as the target baryon with the opposite 
parity, since the pseudoscalar NG boson is bound in the $s$-wave 
interaction. Let us take the ground state octet ${}^2\bm{8}$ and decuplet 
$^4\bm{10}$ as the target baryons. Then the spin-flavor quantum numbers of 
the hadron-NG boson bound states are ${}^2\bm{1}, {}^2\bm{8}, {}^2\bm{8}$ 
from the octet target and ${}^4\bm{8}, {}^4\bm{10} $ from the decuplet 
target, as shown in Table~\ref{tbl:binding}. Comparing the bound states 
obtained in the chiral unitary approach and the first excited states 
belonging to $\bm{70}$ in the quark model, we find that the $^4\bm{1}$ bound
state is absent in our approach but the $^4\bm{10}$ state is present 
instead. It is interesting to see that, in the quark models, the $^4\bm{1}$ 
state is assigned as $\Lambda(1520)$~\cite{Isgur:1978xj}, while in the 
chiral unitary approach it is reproduced as a $^4\bm{8}$ dominant 
state~\cite{Sarkar:2004jh,Roca:2006sz}. Therefore, examination of the 
property of the $\Lambda(1520)$ resonance will provide further understanding
of the baryon spectroscopy, for instance, through the coupling to the vector
mesons~\cite{Hyodo:2006uw}.

\subsection{Critical coupling strength}

\begin{figure}[tbp]
\centerline{\includegraphics[width=0.5\textwidth]{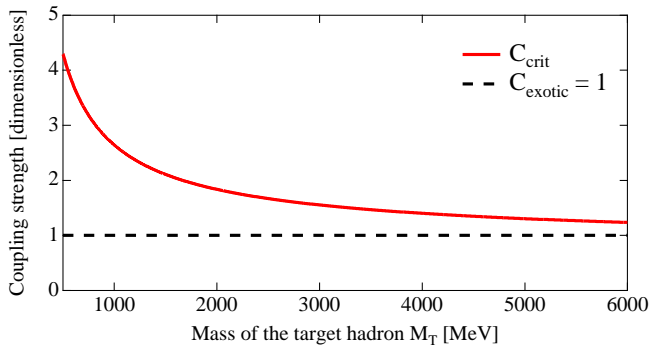}}
\caption{(color online). Critical coupling strength $C_{\text{crit}}$ for 
$m=368$ MeV (solid line). The universal strength of the attraction in exotic
channels $C_{\text{exotic}}=1$ is shown by the dashed line.}
\label{fig:critical}
\end{figure}

As studied in Sec.~\ref{subsec:Bound}, we find the critical attractive 
strength $C_{\text{crit}}$ to generate a bound state. We plot 
$C_{\text{crit}}$ which is evaluated by Eq.~\eqref{eq:WTcritical} as a
function of $M_T$ in Fig.~\ref{fig:critical}, where we employ the meson 
decay constant $f=93$ MeV and the meson mass $m=368$ MeV which corresponds 
to the 
averaged mass of the octet mesons ($\pi$, $K$, $\eta$). We also plot 
$C_{\text{exotic}}=1$, which is the universal strength of the possible 
attraction in exotic channels. It is clear that the attraction 
$C_{\text{exotic}}=1$ is not enough to bind the two-body system for the 
target mass $M<6$ GeV, where all the hadronic target states we consider lie.

Let us examine the robustness of the conclusion. As seen in 
Fig.~\ref{fig:critical} the critical coupling $C_{\text{crit}}$ is 
monotonically decreasing as we increase $M_T$. Therefore it will become 
smaller than $C_{\alpha,T}=1$ at sufficiently large $M_T$. Quantitatively, 
in order to have a bound state for the exotic channel with 
$C_{\text{exotic}}=1$, the mass of the target hadron $M_T$ should be larger 
than about 14 GeV for $m=368$ MeV and $f=93$ MeV. Exotic hadrons can exist 
as bound states of the NG boson-hadron system if a stable hadron (in 
$\bm{3}$ representation, for instance) exists in this energy region. So far 
no hadronic states have been reported.

\begin{figure}[tbp]
\centerline{\includegraphics[width=0.5\textwidth]{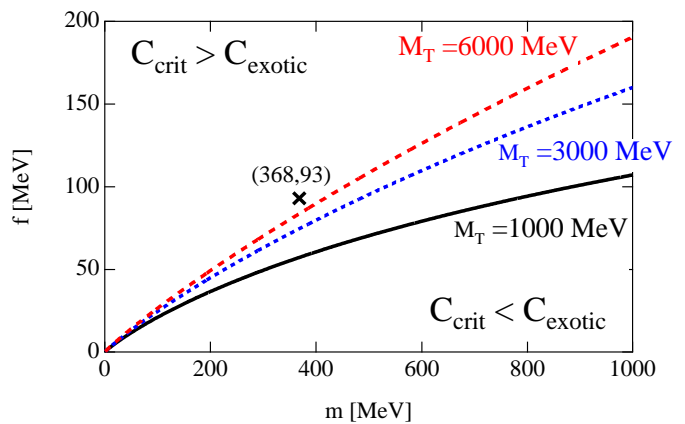}}
\caption{(color online). Lines for $C_{\text{crit}}=1$ in $(m,f)$ plane, 
with $M_T=1000$ MeV (solid line), $M_T=3000$ MeV (dotted line), and 
$M_T=6000$ MeV (dashed line). In the region below (above) the lines, 
$C_{\text{crit}}<1$ ($C_{\text{crit}}>1$). The point of $(m,f)=(368$, $93$) 
MeV is indicated by the cross.}
\label{fig:criticalregion}
\end{figure}

Next we study the dependence of $C_{\text{crit}}$ on $m$ and $f$. It follows
from Eq.~\eqref{eq:WTcritical} that, as the decay constant $f$ is increased,
the critical coupling $C_{\text{crit}}$ also increases. The dependence on 
$m$ is essentially determined by the prefactor $1/m$ and the critical 
coupling $C_{\text{crit}}$ becomes smaller as we increase $m$, since the 
dependence of $G$ on $m$ is not so strong. In Fig.~\ref{fig:criticalregion},
we show the $C_{\text{crit}}=C_{\text{exotic}}$ lines in $m$-$f$ plane with 
three different values of $M_T=1000$, $3000$, and $6000$ MeV. A parameter 
choice below the $C_{\text{crit}}=C_{\text{exotic}}$ lines provides 
$C_{\text{crit}}<1$, so that the exotic states can be bound. Above these 
liens, the attractive interaction in the exotic channels is not strong
enough to generate a bound state.

In Fig.~\ref{fig:criticalregion}, we indicate the point $(m,f)=(368$, $93$) 
MeV by the cross. From this figure, we see that the NG boson could be bound
in the exotic channel, if we use a heavier mass for the NG boson. For 
instance, with $m=$ 500 MeV an exotic bound state appears with $M_{T}\simeq$
2500 MeV as shown in Ref.~\cite{Lutz:2003jw}. One should however note that,
to utilize the low energy theorem of the chiral symmetry, the mass of the NG
boson should not be large. On the other hand, for a smaller NG boson mass, 
the critical strength $C_{\text{crit}}$ becomes larger. Indeed, with 
$m=m_{\pi}$, the critical strength $C_{\text{crit}}\sim 10$ for the mass of 
$N$ and $\Delta$. Using Eq.~\eqref{eq:WTintSU2}, we can show that the 
strongest attraction is $C_{1/2,1/2}=2$ for $T=N$ and $C_{1/2,3/2}=5$ for 
$T=\Delta$. Therefore, attractions found in SU(2) meson-baryon scatterings 
do not generate the bound states, which is consistent with experimental 
observation.

In summary, we show that the critical coupling strength $C_{\text{crit}}$
is larger than the attractive interaction found in exotic channels, with the
physically known values of parameters $M_T$, $m$, and $f$. The critical 
value $C_{\text{crit}}$ could be smaller than $C_{\text{exotic}}$ if the 
mass of the target hadron $M_T$ is sufficiently heavy, or the mass of the NG
boson $m$ is large, or the decay constant of the NG boson $f$ is small.

\section{Summary and discussions}\label{sec:summary}

In this paper, we have studied the $s$-wave bound states in the NG 
boson-hadron scattering described by the chiral unitary approach in flavor 
SU(3) limit. We have studied the group theoretical structure of the 
Weinberg-Tomozawa interaction, which is independent thanks to the chiral 
symmetry. The general expression of the coupling strengths of the WT 
interaction $C_{\alpha,T}$ have been derived. Considering the nonexotic 
hadrons as the target hadrons, we have found an attraction in several exotic
channels with the strength $C_{\text{exotic}}=1$. More generally, based on 
the group theoretical argument, we have shown that the coupling strengths of
the attractive interaction for the channel raising the exoticness are 
universally $C_{\text{exotic}}=1$, which is the smallest value in the WT 
interaction.

We have proved that the weak attraction $C_{\text{exotic}}=1$ of the exotic 
channel in the WT term is not enough to generate the bound state of the 
experimentally observed hadrons and the NG boson. For proof, we have derived
the critical strength of attraction to generate a bound state 
$C_{\text{crit}}$, based on the general principles of the scattering theory.
The critical strength is expressed in terms of the mass of the target hadron
and the mass and decay constant of the NG boson. Studying the dependence of 
the critical strength on these parameters, we find that the exotic bound 
state could be generated by the WT term for larger $M_T$, larger $m$, and 
smaller $f$. We find, however, that these parameters for experimentally
known hadrons do not satisfy $C_{\text{crit}}<C_{\text{exotic}}$. In other 
words, the strength of attraction in the exotic channels is very weak and 
smaller than the critical value, for the experimentally known particles.

We have also examined the large-$N_c$ behavior of the coupling strengths, 
which shows the nontrivial $N_c$ dependence. Because of the $N_c$ dependence
of the coupling strengths, all amplitudes calculated by the WT term for the 
exotic channels become repulsive or vanish in the large-$N_c$ limit, 
including attractive ones at $N_c=3$. The analysis of the critical coupling,
together with the large-$N_c$ behavior of the WT interaction, shows the 
difficulty of generating the exotic state by the chiral interaction. 

In the present approach, exotic hadrons are treated as quasibound states of
the NG boson and a target hadron on the same footing with the nonexotic 
resonances. We have shown that the attractive interaction of the WT term is 
not enough to bind the system. Considering the fact that a certain number of
known resonances have been properly generated by the chiral unitary 
approaches, our conclusion on the exotic states should be of great 
relevance. It should be noted, however, that the present analysis does not 
exclude the existence of the exotic states formed by other mechanisms, for 
instance the genuine quark states, because the WT interaction is one 
specific mechanism to generate states dynamically. 

Apart from the existence of the genuine quark state, one may be cautious of 
the following uncertainties in the analysis; the flavor SU(3) symmetry is 
badly broken in nature, where it is known that the substantial breaking 
effect is of about 20\%. We implicitly assumed that the target particle is 
stable, although several states such as those in $\bm{10}$ can decay via the
strong interactions. In recent studies of the chiral unitary 
approach~\cite{Borasoy:2004kk,Borasoy:2005ie,Oller:2005ig,Oller:2006jw,
Oller:2006yh}, an important role played by the higher order terms of the 
chiral Lagrangian has been discussed, while we only take into account the 
leading order term. The channels which are zero or repulsive at the leading 
order can become attractive when the higher order terms are included. 
Indeed, some resonances can be generated by the effect of the higher order 
terms, which are absent in the leading order 
calculations~\cite{Oller:2006jw}.

Despite these caveats, the present formulation in the SU(3) limit is 
sufficient to discuss the qualitative features of the dynamically generated 
states. For instance, two poles corresponding to the $\Lambda(1405)$ 
resonance are found with and without the SU(3) breaking effects or the 
higher order terms~\cite{Jido:2003cb,Hyodo:2003qa,Borasoy:2005ie,
Oller:2006yh}. This fact indicates that the qualitative feature of the
result is independent of the detailed construction of the model, and that 
the essential structure is determined by the WT term, which is common for 
all approaches. To perform a more quantitative analysis, corrections from 
the above-mentioned effects should be taken into account. This can be 
performed systematically, for instance, based on the chiral perturbation 
theory. For future perspective, it is important to include the SU(3) 
breaking effects to draw more quantitative conclusion. The bound states in 
the SU(3) limit were found to become resonances when the SU(3) breaking 
effect was taken into account. The mechanism for how the bound states 
acquire the width should be clarified, in
order to connect the results in the SU(3) limit to the physical world. These
will be summarized elsewhere.

\begin{acknowledgments}
    The authors are grateful to Professor M.\ Oka for helpful discussions. 
    We also thank Professor V.\ Kopeliovich for useful comments on 
    exoticness. T.~H. thanks the Japan Society for the Promotion of Science 
    (JSPS) for financial support.  This work is supported in part by the 
    Grant for Scientific Research (No.\ 17959600, No.\ 18042001, and 
    No.\ 16540252) and by Grant-in-Aid for the 21st Century COE ``Center for
    Diversity and Universality in Physics'' from the Ministry of Education, 
    Culture, Sports, Science and Technology (MEXT) of Japan.
\end{acknowledgments}

\appendix

\section{Exoticness}\label{sec:Exoticness}

In this section, we derive the exoticness quantum number 
$E$~\cite{Kopeliovich:1990ez,Diakonov:2003ei,Kopeliovich:2003he,
Jenkins:2004tm}. The exoticness expresses the minimal number of the valence 
quark-antiquark pairs to compose the flavor multiplet $[p,q]$ with a given 
baryon number~$B$. The baryon number $B$ here counts only the $u$, $d$, and
$s$ quarks, so that  $B$ can be fractional for the hadrons with the heavy 
quarks. Here we obtain the formula of $E$ for arbitrary $B$ and $[p,q]$ for 
the first time.

Let us consider a hadron with the baryon number $B$ belonging to the $[p,q]$
representation in the flavor SU(3) group, where $p$ and $q$ are nonnegative
integers. From the triality of the $[p,q]$ representation, $p+2q$ is 
congruent to $3B$ modulo 3:
\begin{equation}
    p+2q\equiv 3B \pmod{3} .
    \label{eq:triality} 
\end{equation}
This ensures that the exoticness quantum numbers are integers. We denote the
numbers of the light valence quarks and antiquarks in the hadron as $n_{q}$ 
and $n_{\bar q}$, respectively. Then the baryon number and the exoticness 
are given by 
\begin{eqnarray}
    B&=& \frac{n_{q} - n_{\bar q}}{3} , \label{eq:Bnum} \\
    E&=& \min (n_{q},n_{\bar q}) .\label{eq:Edef}
\end{eqnarray}
The exoticness is given by the antiquark number for the hadron with the 
positive baryon number ($n_{q} > n_{\bar q}$) and the quark number for the 
negative baryon number ($n_{q} < n_{\bar q}$). We have an exception for the
light mesons ($B=0$). Their exoticness is given by
\begin{equation}
   E = n_{q}-1 = n_{\bar q} - 1 \quad \text{for light meson.}
   \label{eq:Emes}
\end{equation}
The valence quark contents $ (n_{q},n_{\bar q})$ of the hadron are uniquely 
determined by the baryon number $B$ and the representation $[p,q]$. 
Therefore, once we know the quark contents $ (n_{q},n_{\bar q})$ of the 
hadron, we obtain the exoticness $E$ through Eqs.~\eqref{eq:Edef} and 
\eqref{eq:Emes}. In the following, we express $n_{q}$ and $n_{\bar{q}}$ in 
terms of $B$, $p$, and $q$.

Let us consider the Young tableau of the representation $[p,q]$, which is 
expressed diagrammatically by
\setlength{\unitlength}{12pt}
\begin{equation}
    \begin{picture}(11,3) 
	\put(0.5,1.5){\framebox(1,1){}}
	\put(0.5,0.5){\framebox(1,1){}}
	\put(1.5,1.5){\framebox(1,1){}}
	\put(1.5,0.5){\framebox(1,1){}}
	\put(3,1.5){$\dots$}
	\put(4.5,1.5){\framebox(1,1){}}
	\put(4.5,0.5){\framebox(1,1){}}
	\put(5.5,1.5){\framebox(1,1){}}
	\put(6.5,1.5){\framebox(1,1){}}
	\put(8,2){$\dots$}
	\put(9.5,1.5){\framebox(1,1){}}
	\put(0.5,3){$\gets$}
	\put(2.5,3){$q$}
	\put(4.5,3){$\to$}
	\put(5.5,3){$\gets$}
	\put(7.5,3){$p$}
	\put(9.5,3){$\to$}
    \end{picture} .
    \nonumber
\end{equation}
In order to know the valence quark contents of the hadron with $B$ and 
$[p,q]$, we assign these boxes to quarks and antiquarks so as to give the 
baryon number $B$ of the hadron. the quark is denoted by one box, while the
antiquark is expressed by two boxes in an antisymmetric combination (two 
boxes in different rows). The baryon number which can be expressed by the 
boxes is limited within the following range:
\begin{equation}
    \llim \leq B \leq \ulim ,
    \label{eq:Brange}
\end{equation}
with 
\begin{equation}
   \llim \equiv \frac{p-q}{3}, \quad \ulim \equiv \frac{p+2q}{3} .
    \label{eq:defnu}
\end{equation}
The lower limit $\llim$ is given by maximizing the number of antiquarks,
while the upper limit $\ulim$ corresponds to the assignment of all the boxes
for quarks. There are two cases: (i) the present boxes are enough to obtain 
the baryon number, that is, the baryon number of the hadron is within the 
range~\eqref{eq:Brange}; (ii) some extra boxes with the flavor singlet 
combination are necessary to express the baryon number. The addition of the 
extra boxes with the flavor singlet combination can be done without changing
the representation~$[p,q]$. 

For case (i), since the number of the boxes in the Young tableau is 
just enough to give the quark contents of the hadron, the total number of 
the boxes, $p+2q$, is given by
\begin{equation}
    p+2q = n_{q}+2n_{\bar{q}}.
    \label{eq:number1}
\end{equation}
Combining Eqs.~\eqref{eq:Bnum} and \eqref{eq:number1}, we obtain the quark 
contents in terms of $B$, $p$, and $q$ as
\begin{equation}
    n_{q}=\frac{p+2q}{3}+2B ,
    \quad 
    n_{\bar{q}}=\frac{p+2q}{3}-B   .
    \nonumber
\end{equation}

For case (ii), the baryon number cannot be obtained from the existing boxes,
so we add some sets of three boxes forming the flavor singlet 
(totally antisymmetric) combination with nonzero baryon number: 
\setlength{\unitlength}{12pt}
\begin{equation}
    \begin{picture}(16,3) 
	\put(0.5, 1.5){\framebox(1,1){}}
	\put(0.5, 0.5){\framebox(1,1){}}
	\put(0.5, -0.5){\framebox(1,1){}}
	\put(1.5, 1.5){\framebox(1,1){}}
	\put(1.5, 0.5){\framebox(1,1){}}
	\put(1.5, -0.5){\framebox(1,1){}}
	\put(3,1.5){$\dots$}
	\put(4.5, 1.5){\framebox(1,1){}}
	\put(4.5, 0.5){\framebox(1,1){}}
	\put(4.5, -0.5){\framebox(1,1){}}
	\put(5.5,1.5){\framebox(1,1){}}
	\put(5.5,0.5){\framebox(1,1){}}
	\put(6.5,1.5){\framebox(1,1){}}
	\put(6.5,0.5){\framebox(1,1){}}
	\put(8,1.5){$\dots$}
	\put(9.5,1.5){\framebox(1,1){}}
	\put(9.5,0.5){\framebox(1,1){}}
	\put(10.5,1.5){\framebox(1,1){}}
	\put(11.5,1.5){\framebox(1,1){}}
	\put(13,2){$\dots$}
	\put(14.5,1.5){\framebox(1,1){}}
	\put(5.5,3){$\gets$}
	\put(7.5,3){$q$}
	\put(9.5,3){$\to$}
	\put(10.5,3){$\gets$}
	\put(12.5,3){$p$}
	\put(14.5,3){$\to$}
    \end{picture} .
    \nonumber
\end{equation}
Note that the addition of the flavor singlet with zero baryon number is 
meaningless, since it supplies just sea quark pairs. An example of case (ii)
is the flavor singlet hadron with $B=1$. Since the Young tableau of the
flavor singlet $[0,0]$ has no box, the baryon number $B=1$ cannot be 
achieved by the present box. Thus we add one set of three boxes with the
flavor singlet and assign quarks to these boxes to give $B=1$. Then we 
obtain the valence quark contents as $n_{q}=3$ and $n_{\bar q}=0$ for the 
flavor singlet hadron with $B=1$.

Case (ii) can be further classified into two cases:
\begin{equation}
B < \llim \quad \text{or} \quad  \ulim < B \ .   \label{eq:caseii}
\end{equation}
For the case $\ulim < B $, the baryon number $B$ exceeds the upper limit 
$\ulim$. Recalling that $\ulim$ is obtained so as to assign all the boxes to
quarks, that is, the quark number is $p+2q$ and the antiquark number is 
zero, we see that the boxes for quarks are not enough to express the baryon 
number. To fill the deficit, we add $B-\ulim$ sets of three boxes in one 
column forming the flavor singlet and assign quarks to the boxes. The 
addition of one set of boxes changes the baryon number by $+1$. Hence, we 
obtain
\begin{equation}
    n_{q} = p+2q + 3(B-\ulim) = 3B,\quad n_{\bar{q}}= 0 .
    \nonumber
\end{equation}
For the case $B < \llim$, the baryon number $B$ is less than the lower limit
$\llim$. In the lower limit, the antiquarks are maximally assigned to the 
boxes, that is, the quark number is $p$ and the antiquark number is $q$. 
Since $B < \llim$ means a lack of boxes for antiquarks to express the 
baryon number, we add $\llim-B$ sets of the flavor singlet combination and
maximally assign antiquarks to the whole boxes. The addition of $\llim-B$
sets of boxes changes the total number by $-1$. Hence we obtain 
\begin{align}
    n_{q} =& p-(\llim-B) = \ulim+\llim+B, \nonumber \\
    n_{\bar{q}} =& q+2(\llim-B) =\ulim+\llim-2B .
    \nonumber
\end{align}

Summarizing all the cases, we obtain the valence quark contents of the
hadron with the baryon number $B$ belonging to the SU(3) representation
$[p,q]$:
\begin{align}
   n_{q} &=
      \begin{cases} 
         3B & \text{for} \quad \ulim \leq B ,\\
        \ulim + 2B &\text{for} \quad  \llim \leq B \leq  \ulim , \\
         \ulim+\llim + B & \text{for}\quad B\leq \llim ,
      \end{cases} \label{eq:qnum} \\
     n_{\bar q} &=
      \begin{cases} 
         0 & \text{for} \quad \ulim \leq B , \\
         \ulim - B &\text{for} \quad \llim\leq B \leq  \ulim ,\\
        \ulim+\llim - 2B & \text{for} \quad B \leq \llim ,
      \end{cases} \label{eq:aqnum} 
\end{align}
where $\llim$ and $\ulim$ are defined in Eq.~\eqref{eq:defnu}. Substituting 
Eqs.~\eqref{eq:qnum} and \eqref{eq:aqnum} into Eqs.~\eqref{eq:Edef} and 
\eqref{eq:Emes}, we obtain the exoticness for the hadron with $B$ and
$[p,q]$. 

In the present study, we are interested in $B>0$. In this case, the 
exoticness $E=n_{\bar{q}}$ is given by
\begin{equation}
    E=
    \begin{cases}
        0 & \text{for} \quad \epsilon \leq 0 , \\
	\epsilon   & \text{for}\quad  \epsilon \geq 0, \ \nu \leq 0 ,  \\
	\epsilon +\nu   & \text{for}\quad \nu \geq 0 ,
	\end{cases} 
    \nonumber
\end{equation}
with $ \epsilon \equiv \ulim -B$ and $ \nu \equiv \llim -B$. 
This is equivalently expressed as
\begin{equation}
    E=\epsilon\theta(\epsilon)+\nu\theta(\nu) ,
    \nonumber
\end{equation}
with the step function $\theta(x)$. 

The exoticness quantum number was introduced in the context of the chiral 
soliton models~\cite{Kopeliovich:1990ez,Diakonov:2003ei}, but without 
considering the case~\eqref{eq:caseii}. The insufficiency of the definition 
was realized~\cite{Kopeliovich:2003he} and later the definition was 
generalized to give correct values for $B<\beta_-$~\cite{Jenkins:2004tm}, 
while the definition of $B>\beta_+$ states was still not properly treated. 
This is enough for the states in the excited spectrum of soliton rotation, 
since there is no state with $B>\beta_+$. In our definition, the states with
$B>\beta_+$ can be correctly treated, and the generalization to the 
arbitrary baryon number is performed.

The exoticness derived above can be extended to the baryon with arbitrary 
$N_c$. The minimal contents of the baryon is $N_c$ quarks. The baryon number
for a hadron with $n_Q$ heavy quarks is given by $B=1-n_Q/N_c$. From this 
and the triality of the flavor SU(3), we have, instead of 
Eqs.~\eqref{eq:triality} and \eqref{eq:Bnum},
\begin{align}
    p+2q&\equiv N_cB \pmod{3} ,\quad
    B= \frac{n_{q} - n_{\bar q}}{N_c} .
    \nonumber
\end{align}
Following the same argument as before with the replacement $B\to 
\frac{N_c}{3} B$, we obtain the numbers of quarks and antiquarks for 
arbitrary $N_c$,
\begin{align}
   n_{q} &=
      \begin{cases} 
        N_cB & \text{for} \quad \ulim \leq \frac{N_c}{3}B , \\
        \ulim + \frac{2N_c}{3}B &
	\text{for} \quad  \llim \leq \frac{N_c}{3}B \leq  \ulim ,\\
        \ulim+\llim + \frac{N_c}{3}B & \text{for}\quad 
	\frac{N_c}{3}B\leq \llim ,
      \end{cases} \nonumber \\
     n_{\bar q} &=
      \begin{cases} 
         0 & \text{for} \quad \ulim \leq \frac{N_c}{3}B ,\\
         \ulim - \frac{N_c}{3}B
	 &\text{for} \quad \llim\leq \frac{N_c}{3}B \leq  \ulim ,\\
         \ulim+\llim - \frac{2N_c}{3}B & \text{for} \quad
	 \frac{N_c}{3}B \leq \llim .
      \end{cases} \nonumber 
\end{align}
From Eq.~\eqref{eq:Edef}, the exoticness of a baryon is given by 
$E=n_{\bar{q}}$.

\section{Delta function potential}\label{sec:Delta}

Let us consider the $d$-dimensional delta function potential problem in 
quantum mechanics~\cite{Jackiw:1991je}. The Schr\"odinger equation takes on
the form with $\hbar = 1$
\begin{align}
    -\frac{1}{2m} \bm{\nabla}^2 \psi(\bm{r})
    +V(\bm{r})\psi(\bm{r}) &= E\psi(\bm{r}),
    \nonumber
\end{align}
with the reduced mass of the system $m$ and the $d$-dimensional delta 
function potential $V(\bm{r})=-v\delta(\bm{r})$. We consider the attractive
interaction $v>0$. To get an essential feature of the problem, it is 
convenient to consider it in a variational method. Let us take an 
expectation value of the Hamiltonian of the sum of the kinetic and potential
terms with a normalized $s$-wave trial wave function with a size parameter, 
say $b$, as a variational parameter. Then the energy plot as a function of 
$b$ has one minimum for $d = 1$, while it has no such a stable point for 
$d = 3$; it falls into $- \infty$ as $b \to 0$, meaning that the delta 
function potential is too strongly attractive to overcome the kinetic energy
due to the uncertainty principle.

For more quantitative discussions, we consider the wave function explicitly 
which is given in momentum space as
\begin{equation}
    \phi(\bm{p})
    =\frac{v_0 \psi(\bm{0})}{p^2 +\tilde{E}_{b}} ,
    \nonumber
\end{equation}
where $v_0=2mv$ and $\tilde{E}_b=2mE_b\equiv-2mE>0$. Integrating both sides
over $\bm{p}$, we obtain
\begin{align*}
    \frac{1}{v_0}
    &= 
    \frac{1}{(2\pi)^d}
    \int d^d\bm{p}
    \frac{1}{p^2 +\tilde{E}_{b}}  \\
    &=\frac{1}{(2\pi)^d}
    \int d\Omega
    \int dp p^{d-1}
    \frac{1}{p^2 +\tilde{E}_{b}} .
\end{align*}
Here we use $\int d^{d} \bm{p} \, \phi(\bm{p}) = \psi(\bm{0})$. Notice that 
the integral is divergent for $d\geq 2$ so that we need regularization. For
$d=3$, by introducing the cutoff $\Lambda$ in the three-momentum, this 
integration leads to
\begin{align}
    \frac{1}{v_0}
    =& \frac{1}{2\pi^2}
    \left(
    \Lambda -\sqrt{\tilde{E}_{b}}\arctan
    \left[\frac{\Lambda}{\sqrt{\tilde{E}_{b}}}\right]
    \right) .
    \label{eq:boundstate}
\end{align}
The energy of the bound state $E_b$ can be obtained by solving this 
equation, which depends not only on $v$ but also $\Lambda$. In addition, 
we need a condition in order to obtain the solution for 
Eq.~\eqref{eq:boundstate}:
\begin{align}
    v>\frac{\pi^2}{m\Lambda}\equiv v_c .
    \nonumber
\end{align}
This equation determines the critical strength of attraction $v_c$ that can 
provide the bound state, for a given cutoff $\Lambda$. If the attraction is
weaker than $v_c$, no bound state appears.

For the scattering state, the wave function in momentum space $\phi(\bm{p})$
is given by
\begin{align}
    \phi(\bm{p})
    =&(2\pi)^3\delta(\bm{p}-\bm{k})
    +\frac{v_0}{p^2-k^2-i\epsilon}\psi(\bm{0}) ,
    \nonumber
\end{align}
where $k^2=2mE$. Integrating both sides over $\bm{p}$, we obtain the 
scattering amplitude $f(k)$ as
\begin{align}
    f(k)=&
    \frac{v_0\psi(\bm{0})}{2\pi}
    =\frac{1}{2\pi}\left(\frac{1}{v_0}
    -\frac{1}{2\pi^2}
    \left[\Lambda+\frac{k}{2}
    \ln\frac{k-\Lambda}{k+\Lambda}
    \right]
    \right)^{-1}
    \nonumber \\
    \equiv & \frac{1}{2\pi D(k)} . \nonumber
\end{align}
Here we use, for a complex value $k$, 
\begin{equation}
   I =  \int_0^{\Lambda}
    dp\frac{p^2}{p^2-k^2-i\epsilon}
    = \Lambda+\dfrac{k}{2}
       \ln\dfrac{k-\Lambda}{k+\Lambda} .
       \nonumber 
\end{equation}
This integral $I$ with the complex $k$ is obtained through analytic 
continuation of the integral with a real value $k$:
\begin{equation*}
   I = 
    \begin{cases}
       \Lambda+\dfrac{k}{2}
       \left[\ln\left|\dfrac{\Lambda-k}{\Lambda+k}\right|+i\pi
       \right]
       & \text{for} \quad  k <\Lambda\\
       \Lambda+\dfrac{k}{2}
       \ln\left|\dfrac{\Lambda-k}{\Lambda+k}\right|
       & \text{for} \quad  k >\Lambda
    \end{cases} \ .
\end{equation*}

Bound states and resonances appear when $f(k)$ has poles [or equivalently, 
$D(k)=0$] at 
\begin{align}
    &\re [k] = 0, \quad \im [k] >0 \quad \text{for a bound state,} 
    \nonumber \\
    &\re [k] > 0, \quad \im [k] <0 \quad \text{for a resonance state.}
    \label{eq:resonancecond}
\end{align}
Indeed, the condition that $D(k)$ should be zero for pure imaginary $k$ is 
equivalent to Eq.~\eqref{eq:boundstate}. In order to obtain a resonance 
state, the imaginary part of $D(k)$ should necessarily be zero with
$\im [k] <0$, which can  be decomposed as
\begin{align*}
    \im[D(k)] 
    & =  -\frac{1}{2\pi^2}\Biggl(
    \frac{\re[k]}{2}
    \arg\frac{k-\Lambda}{k+\Lambda}
    +\frac{\im[k]}{2}
    \ln\left|\frac{k-\Lambda}{k+\Lambda}\right|
    \Biggr) .
\end{align*}
Now the first term should be positive because the argument is defined in the
range $0\leq \theta <2\pi$. Since $|k+\Lambda|>|k-\Lambda|$,
\begin{equation}
    \ln\left|\frac{k-\Lambda}{k+\Lambda}\right|<0 ,
    \nonumber
\end{equation}
and therefore the second term is also positive. Thus we find $\im[D(k)]<0$ 
for Eq.~\eqref{eq:resonancecond}, and no resonance state appears.

\section{Heavy meson target}\label{sec:mesoncase}

Here we address the formulation of the chiral unitary approach for the case 
with the meson target. The scattering amplitude is given in the same way 
as Eq.~\eqref{eq:ChUamp}:
\begin{equation}
    t=V+VGt=
    \frac{1}{1-VG}V ,
    \label{eq:CUM}
\end{equation}
with the interaction $V$ and the loop function $G$. Denoting $V_B$ and $G_B$
for a baryon target equivalent to the expressions in Sec.~\ref{sec:ChU} and 
$V_M$ and $G_M$ for a meson target given in Ref.~\cite{Kolomeitsev:2003ac}, 
we have
\begin{align}
    V_B
    &=-\frac{1}{2f^2}C\omega ,
    \nonumber \\
    V_M
    &=-\frac{1}{8f^2}C
    \left(
    3s-2M_T^2-2m^2-\frac{(M_T^2-m^2)^2}{s}
    \right)  ,
    \nonumber
\end{align}
where $\sqrt{s}$ is the total energy and $M_T$ and $m$ are the masses of the
target and the NG boson. We have used our convention for the coupling 
strength $C$. The loop function is given by
\begin{align}
    G_B
    &=
    i\int\frac{d^{4}q}{(2\pi)^{4}}
    \frac{2M_T}{(P-q)^{2}-M_T^{2}+i\epsilon}
    \frac{1}{q^{2}-m^{2}+i\epsilon}
    \nonumber ,  \\
    G_M
    &=\frac{G_B}{2M_T}
    \nonumber  .
\end{align}
Note that the dimensions of $G$ and $V$ in the two cases are different, 
reflecting the factor $2M_T$ which comes from the normalization of the 
fermion spinor $\bar{u}u=1$. 

We expand the interaction potentials in terms of $\omega/M_T$, where 
$\omega$ is the NG boson energy. Since we are interested in the low energy 
dynamics of the NG boson, the energy of the NG boson is of order of its 
mass. The target meson mass is taken to be large in comparison with the NG 
boson mass. Since $\sqrt{s}$ also depends on $M_T$, it is better to express 
the amplitudes as functions of the meson energy $\omega$ to take the large 
$M_T$ limit. Expanding the interaction $V_{M}$ with respect to $\omega/M_T$,
we obtain
\begin{equation}
     V_M =  -2 M_T \frac{\omega}{2f^2}C\left(1 + 
    \frac{-m^2+\omega^2}{2\omega^{2}}\frac{\omega}{M_T}\right)
    +\mathcal{O}\left(\left(\frac{\omega}{M_T}\right)^2\right)
\nonumber .
\end{equation}
The leading term agrees with the interaction $V_{B}$, except for the trivial
normalization factor~$2M_T$:
\begin{equation}
     V_B = \frac{V_M}{2M_T} \nonumber .
\end{equation}
This also indicates, from Eq.~\eqref{eq:CUM}, that the scattering amplitudes
also coincide up to ${\cal O}((\omega/M_T))$:
\begin{equation}
    t_B= \frac{t_M}{2M_T}
    +\mathcal{O}(\omega/M_T) .
    \nonumber
\end{equation}
Thus, we have shown that the formulation given in Sec.~\ref{sec:ChU} can be 
applied to the heavy meson target, up to ${\cal O}(\omega/M_T)$ corrections.
The convergence becomes much better, if we include the next to leading order
terms in $V_B$:
\begin{align}
    V^{(2)}_{B}
    =&-\frac{1}{2f^2}C(\sqrt{s}-M_T) 
    \frac{E_T+M_T}{2M_T} .
    \label{eq:WTintApp} 
\end{align}
Here $E_T$ is the on-shell energy of the baryon, and $\sqrt{s}$ is the total
energy in the center-of-mass system. This agrees with the interaction 
$V_{M}$ up to ${\cal O}((\omega/M_T)^{2})$.

\end{document}